\documentclass[12pt]{article}
\usepackage{amsmath,amsthm}
\usepackage{setspace}
\usepackage{graphicx}
\usepackage{enumerate}
\usepackage{natbib}
\usepackage{epsfig}
\usepackage{epstopdf}
\usepackage{color}
\usepackage{authblk}
\usepackage{multirow,booktabs}
\usepackage{multicol}
\usepackage{subcaption}
\usepackage{tabularx}
\usepackage{url,enumerate, amssymb, amsfonts}
\usepackage[colorlinks=true, linkcolor=blue, urlcolor=blue, citecolor=blue, anchorcolor=blue, pdfstartview=FitH, bookmarksdepth=2]{hyperref}

\newtheorem{theorem}{Theorem}

\newtheorem{remark}{Remark}
\allowdisplaybreaks

\makeatletter
\@addtoreset{equation}{section}

\newcommand{\calD}{{\cal D}}

\newcommand{\calC}{{\cal C}}
\newcommand{\calS}{{\cal S}}
\newcommand{\calF}{{\cal F}}

\newcommand{\bM}{\mbox{\bf M}}
\newcommand{\bB}{\mbox{\bf B}}
\newcommand{\bD}{\mbox{\bf D}}
\newcommand{\bZ}{\mbox{\bf Z}}
\newcommand{\1}{\mbox{\bf 1}}
\newcommand{\bg}{\mbox{\bf g}}

\newcommand{\balpha}{\mbox{\boldmath{$\alpha$}}}
\newcommand{\bbeta}{\mbox{\boldmath{$\beta$}}}

\newcommand{\bY}{\mbox{\bf Y}}

\newcommand{\ba}{\mbox{\bf a}}

\newcommand{\bb}{\mbox{\bf b}}

\newcommand{\0}{\bf 0}

\begin{document}

\def\spacingset#1{\renewcommand{\baselinestretch}%
{#1}\small\normalsize} \spacingset{1.5}
\title{\LARGE \bf
A Flexible and Parsimonious Modelling Strategy for Clustered Data Analysis}
\author[1]{Tao Huang}
\author[2]{Youquan Pei}
\author[1]{Jinhong You}
\author[3]{Wenyang Zhang\thanks{The corresponding author, Email: \texttt{wenyang.zhang@york.ac.uk}}}
\affil[1]{\small School of Statistics and Management, Shanghai University of Finance and Economics, CN}
\affil[2]{\small School of  Economics, Shandong University, CN}
\affil[3]{\small Department of Mathematics, University of York, UK}
\date{}
\maketitle

\bigskip
\begin{abstract}
Statistical modelling strategy is the key for success in data analysis.  The
trade-off between flexibility and parsimony plays a vital role in statistical
modelling.  In clustered data analysis, in order to account for the
heterogeneity between the clusters, certain flexibility is necessary in the
modelling, yet parsimony is also needed to guard against the complexity and
account for the homogeneity among the clusters. In this paper, we propose a
flexible and parsimonious modelling strategy for clustered data analysis.  The
strategy strikes a nice balance between flexibility and parsimony, and accounts
for both heterogeneity and homogeneity well among the clusters, which often
come with strong practical meanings. In fact, its usefulness has gone beyond
clustered data analysis, it also sheds promising lights on transfer learning.
An estimation procedure is developed for the unknowns in the resulting
model, and asymptotic properties of the estimators are established.  Intensive
simulation studies are conducted to demonstrate how well the proposed methods
work, and a real data analysis is also presented to illustrate how to apply the
modelling strategy and associated estimation procedure to answer some real
problems arising from real life.
\end{abstract}

\noindent%

\vfill

\newpage
\spacingset{1.9} 
\section{Introduction}
\label{intr0}

The data, which stimulates this paper, come from 29 provinces (for the sake of
convenience, municipalities are also called provinces in this paper) in China.
It includes the daily number of Covid-19 infected cases, of people from Wuhan
to a province, of cured cases, and maximum daily temperature in the province
from $9$ January 2020 to $25$ March 2020, it also includes the population size
of the province.  This dataset is a typical clustered dataset with each individual
of length 77.  We focus on the log ratio, $y_{i,t}$, of the cumulative number
of infected cases in the $i$th province at time point $t$ to that at time point
$t-1$.  We call $y_{i,t}$ the infection ratio at time point $t$ in the
$i$th province.  As cured cases may not have much effect on infection ratio,
we exclude it from our analysis and focus on maximum daily temperature in a
province, denoted by $x_{i,t,1}$, and the number of people from Wuhan to the
province, denoted by $x_{i,t,2}$.  What we are interested in are: how maximum
daily temperature in a province and the number of people from Wuhan to the
province affect the infection ratio in the province?  Whether their impacts
change over time?  If they do, what are the dynamic patterns of the impacts?
Whether the impacts vary over different provinces?  Are there provinces
sharing the same impacts?  If there are, which provinces share the same
impacts?

The varying coefficient models are a powerful tool to explore nonlinear dynamic
patterns of impacts of covariates.  There is much literature about the varying
coefficient models \citep{zl2000,zls2002,liu2014feature,cheng2016peter, cheng2016forward,chu2020feature}, and its application in the analysis of time
series data \citep{guodongli2008,guodongli2011,xia2015,lei2016estimation} and
of clustered data \citep{lkz2015,li2017functional,zhong2019homogeneity,xyz2020,
feng2018varying}.  We start with the standard varying coefficient models, which leads to
\begin{equation}
y_{i,t}
=
b_i + a_{i,1}(u_{i,t}) x_{i,t,1} + a_{i,2}(u_{i,t}) x_{i,t,2} + \epsilon_{i,t},
\quad
u_{i,t} = t/77.
\label{sta0}
\end{equation}
Whilst the dynamic patterns of the impacts of $x_{i,t,1}$ and $x_{i,t,2}$ have
been reflected and formulated by $a_{i,1}(\cdot)$ and $a_{i,2}(\cdot)$, this
modelling assumes the contribution of covariates to response is linear,
functional coefficients though. This may not be realistic, indeed, in Section
\ref{real0}, we will see this assumption is not valid, and the contribution of
covariates is in fact through transformed covariates.  Furthermore, this
modelling has not taken into account the effects of previous infection ratios.
To overcome these problems takes us to the following model
\begin{eqnarray}
y_{i,t}
& = &
b(u_{i,t})
+
\sum\limits_{j=1}^p a_{i,j}(u_{i,t}) g_{i,j}(y_{i,t-j})
+
a_{i,p+1}(u_{i,t}) g_{i,p+1}(x_{i,t,1})
\nonumber
\\
& &
+
a_{i,p+2}(u_{i,t}) g_{i,p+2}(x_{i,t,2})
+
\eta_i + \epsilon_{i,t},
\label{sta1}
\end{eqnarray}
where $b(\cdot)$, $a_{i,k}(\cdot)$s and $g_{i,k}(\cdot)$s are unknown functions,
and $\eta_i$s are random effects.  To answer the question that which provinces
share the same impacts, we assume some of $a_{i,k}(\cdot)$s are the same, and
some of $g_{i,k}(\cdot)$s are the same.  We don't assume which
$a_{i,k}(\cdot)$s or $g_{i,k}(\cdot)$s are the same, and let data identify.
Putting this modelling idea in a more formal and generic form leads to the
following class of semiparametric models for clustered data analysis.

Rather than confining us to the Covid-19 dataset, we assume $y_{i,t}$ is the
observation of the response variable of the $i$th individual at time point $t$,
$(u_{i,t}, \mathbf{X}_{i,t})$ the vector of the corresponding covariates, where
$\mathbf{X}_{i,t}=(x_{i,t,1},\ldots,x_{i,t,q})^{\tau}$ is of dimension $q$, $u_{i,t}$ is a scalar.  We assume
\begin{equation}
y_{i,t}
=
b(u_{i,t})
+
\sum\limits_{j=1}^p a_{i,j}(u_{i,t}) g_{i,j}(y_{i,t-j})
+
\sum\limits_{l=1}^q a_{i,p+l}(u_{i,t}) g_{i,p+l}(x_{i,t,l})
+
\eta_i
+
\epsilon_{i,t}
\label{model}
\end{equation}
when $t=p+1, \ \cdots, \ T_i$ \footnote{We allow the time series length $T_{i}$
to be heterogeneous across $i$, which corresponding to unbalanced clustered
data.}, $i=1, \ \cdots, \ n$.  $x_{i,t,l}$ is the $l$th
component of $\mathbf{X}_{i,t}$, $\eta_i$ a random effect with mean $0$ and variance
$\sigma_{\eta}^2$.  $\epsilon_{i,t}$ is a random error with mean $0$, variance
$\sigma^2$.  $b(\cdot)$, $a_{i,j}(\cdot)$s, and $g_{i,j}(\cdot)$s are all
unknown functions to be estimated.

Apparently, (\ref{model}) is not identifiable. To make it identifiable, we
assume
\begin{equation}
E\{a_{i,j}(u_{i,t})\} = 1,
\
E\{g_{i,j}(y_{i,t-j})\} = 0,
\
E\{a_{i,p+l}(u_{i,t})\} = 1,
\
E\{g_{i,p+l}(x_{i,t,l})\} = 0,
\label{con}
\end{equation}
for $i=1, \dots, n$, $j=1, \dots, p$ and $l=1, \dots,  q$.

We also assume
\begin{equation}
a_{i,j}(\cdot)
=
\left\{
\begin{array}{cc}
\alpha_1(\cdot) & \mbox{if } (i,j) \in \calD_1
\\
\vdots & \vdots
\\
\alpha_H(\cdot), & \mbox{if } (i,j) \in \calD_H
\end{array}
\right.
\quad
g_{i,j}(\cdot)
=
\left\{
\begin{array}{cc}
\beta_1(\cdot) & \mbox{if } (i,j) \in \Delta_1
\\
\vdots & \vdots
\\
\beta_m(\cdot) & \mbox{if } (i,j) \in \Delta_m
\end{array}
\right.
\label{hom}
\end{equation}
where $\alpha_k(\cdot)$'s and $\beta_k(\cdot)$'s are unknown functions to be
estimated, and $\{\calD_1, \dots, \calD_H\}$ is an unknown partition of
$$
\{(i, j): \ i=1, \ \cdots, \ n; \ j=1, \ \cdots, \ p+q\},
$$
and so is $\{\Delta_1, \ \cdots, \Delta_m\}$.   Except for the practical implication
mentioned before, condition~(\ref{hom}) also serves, in statistical modelling, to make model~(\ref{model}) more parsimonious.
Model~(\ref{model}) under conditions~(\ref{con}) and~(\ref{hom}) is the
model we are going to address in this paper.

We would like to stress that~(\ref{model}) is a large class of semiparametric
models, which include nonparametric autoregressive models \citep{huang2004identification, syzz2014,lei2016estimation,
sun2016functional,kalli2018bayesian}, varying coefficient models and additive models
\citep{wood2015generalized, chen2018error, sang2020estimation}.

Condition~(\ref{hom}) can also be viewed as a latent structure.  Latent structure identification is a very useful statistical modelling idea and has been
widely used in the analysis of cross sectional data
\citep{ke2015homogeneity,fan2017change,ren2019tuning,wu2020statistical,
fan2022simple,yuan2022testing}, of time series data \citep{bahadori2015functional,klz2020} and of clustered data \citep{bonhomme2015grouped,klz2016,su2016identifying,chen2018estimating,li2019subgroup,xyz2020,lqz2021,xiao2021homogeneity,
guo2022homogeneity,pei2022network,zhu2022simultaneous}.

In this paper, we make contributions on three fronts.  Firstly, we propose a
flexible and parsimonious modelling strategy for clustered data analysis, which
results in a large class of semiparametric models, embedded with latent
structures.  It takes dynamic nonparametric models, varying coefficient models
and additive models as its special cases.  Secondly, we develop a procedure,
which is easy to implement by the proposed computational algorithm, to
identify the latent structures and estimate cluster-specific functions.
Thirdly, we demonstrate the advantage of the proposed methodology by asymptotic
theory and empirical analysis.

The rest of the paper is organised as follows. We describe our estimation
procedure and the computational algorithm to implement it in
Section~\ref{est0}.  In Section~\ref{asy0}, we present the asymptotic
properties of the estimators resulted from either the proposed estimation,
overfitting, or underfitting.  Overfitting and underfitting will be defined at
the beginning of Section~\ref{asy0}.  Intensive simulation studies are
conducted in Section~\ref{simu0} to demonstrate how well the proposed
estimation procedure works and the risk of ignoring homogeneity or
heterogeneity among the individuals in a clustered dataset.  In
Section~\ref{real0}, we apply the proposed models and estimation procedure to
the Covid-19 dataset, and explore how maximum daily temperature in a province
and the number of people from Wuhan to the province affect the infection ratio
in the province.  We will identify which provinces in China share the same
impacts, and find out the dynamic patterns of the impacts.  Throughout this
paper, a superscript $\tau$ indicates the transpose of a vector or a matrix.
 $\stackrel{D}{\longrightarrow}$ indicates convergence in a distribution.

\section{Estimation procedure}
\label{est0}

\subsection{Estimation method}
\label{met0}

The proposed estimation procedure consists of three stages: initial estimation,
structure identification, and final estimation. For the sake of convenience in the structure identification in our estimation
procedure, without loss of generality, we assume the range of each covariate
involved in (\ref{model}) and of $y_{i,t}$ is $[0, \ 1]$.

We apply B-spline decomposition, with knots being equally placed, to deal with
the unknown functions in (\ref{model}).  The number of knots used for the final
estimation is larger than that for the initial estimation due to the common
structure identified being used, therefore, the basis functions for the final
estimation are different to that for the initial estimation.  Specifically, we
use $B_{\ell}(\cdot)$, $\ell=1, \ \cdots, \ K$, to denote the B-spline basis
functions used for the final estimation, $B_{\ell}^{(0)}(\cdot)$, $\ell=1, \
\cdots, \ K_0$, for the initial estimation.  Let
$$
\bar{B}_{i,u,\ell}
=
\frac{1}{T_i} \sum\limits_{t=1}^{T_i} B_{\ell}(u_{i,t}),
\quad
\bar{B}_{i,y,j, \ell}
=
\frac{1}{T_i-j} \sum\limits_{t=1+j}^{T_i} B_{\ell}(y_{i,t-j}),
$$
$$
\bar{B}_{i,x,l,\ell}
=
\frac{1}{T_i} \sum\limits_{t=1}^{T_i} B_{\ell}(x_{i,t,l}).
\quad
\bB_{i,0,t}
=
\left(
B_1(u_{i,t}), \ \cdots, \ B_K(u_{i,t})
\right)^{\tau},
$$
$$
\bB_{i,u,t}
=
\left(
B_1(u_{i,t})-\bar{B}_{i,u,1}, \ \cdots, \ B_K(u_{i,t})-\bar{B}_{i,u,K}
\right)^{\tau},
$$
$$
\bB_{i,y,j,t}
=
\left(
B_1(y_{i,t-j})-\bar{B}_{i,y,j,1}, \ \cdots, \ B_K(y_{i,t-j})-\bar{B}_{i,y,j,K}
\right)^{\tau},
$$
$$
\bB_{i,x,l,t}
=
\left(
B_1(x_{i,t,l})-\bar{B}_{i,x,l,1}, \ \cdots, \ B_K(x_{i,t,l})-\bar{B}_{i,x,l, K}
\right)^{\tau}.
$$
Under condition~(\ref{con}), in the final estimation,
\begin{equation}
\begin{array}{c}
b(u_{i,t})
\approx
\bB_{i,0,t}^{\tau} \bb,
\quad
a_{i,j}(u_{i,t})
\approx
1 + \bB_{i,u,t}^{\tau} \ba_{i,j},
\quad
a_{i,p+l}(u_{i,t})
\approx
1 + \bB_{i,u,t}^{\tau} \ba_{i,p+l}
\\
g_{i,j}(y_{i,t-j})
\approx
\bB_{i,y,j,t}^{\tau} \bg_{i,j},
\quad
g_{i,p+l}(x_{i,t,l})
\approx
\bB_{i,x,l,t}^{\tau} \bg_{i,p+l},
\end{array}
\label{app1}
\end{equation}
where $i=1, \dots, n$, $j=1, \dots, p$ and $l=1, \dots, q$.  We use superscript
$(0)$ to denote the counterpart of this notation in the initial estimation,
e.g.\ $\bB_{i,0,t}^{(0)}$ is the counterpart of $\bB_{i,0,t}$.  In the initial
estimation,
\begin{equation}
\begin{array}{c}
b(u_{i,t})
\approx
\bB_{i,0,t}^{(0)\tau} \bb^{(0)},
\quad
a_{i,j}(u_{i,t})
\approx
1 + \bB_{i,u,t}^{(0)\tau} \ba_{i,j}^{(0)},
\quad
a_{i,p+l}(u_{i,t})
\approx
1 + \bB_{i,u,t}^{(0)\tau} \ba_{i,p+l}^{(0)}
\\
g_{i,j}(y_{i,t-j})
\approx
\bB_{i,y,j,t}^{(0)\tau} \bg_{i,j}^{(0)},
\quad
g_{i,p+l}(x_{i,t,l})
\approx
\bB_{i,x,l,t}^{(0)\tau} \bg_{i,p+l}^{(0)}.
\end{array}
\label{app0}
\end{equation}

\subsubsection{Initial estimation}
\label{ini0}

We treat all unknown functions in (\ref{model}) as different functions to
estimate.  Applying the least squares estimation and the approximations in
(\ref{app0}), we have the following objective function
\begin{equation}
\begin{array}{r}
\sum\limits_{t=p+1}^{T_i}
\left\{
y_{i,t}
-
\bB_{i,0,t}^{(0)\tau} \bb_i^{(0)}
-
\sum\limits_{j=1}^p
\left(
1 + \bB_{i,u,t}^{(0)\tau} \ba_{i,j}^{(0)}
\right)
\bB_{i,y,j,t}^{(0)\tau} \bg_{i,j}^{(0)}
-
\right.
\\
\left.
\sum\limits_{l=1}^q
\left(
1 + \bB_{i,u,t}^{(0)\tau} \ba_{i,p+l}^{(0)}
\right)
\bB_{i,x,l,t}^{(0)\tau} \bg_{i,p+l}^{(0)}
\right\}^2.
\end{array}
\label{ls1}
\end{equation}
We minimise~(\ref{ls1}) with respect to $\bb_i^{(0)}$, $\ba_{i,j}^{(0)}$s,
$\bg_{i,j}^{(0)}$s, $\ba_{i,p+l}^{(0)}$s and $\bg_{i,p+l}^{(0)}$s, and denote
the minimisers as $\tilde{\bb}_i^{(0)}$, $\tilde{\ba}_{i,j}^{(0)}$s,
$\tilde{\bg}_{i,j}^{(0)}$s, $\tilde{\ba}_{i,p+l}^{(0)}$s and
$\tilde{\bg}_{i,p+l}^{(0)}$s.  We use
$$
\tilde{b}_i(\cdot)
=
\left(
B_1^{(0)}(\cdot), \ \cdots, \ B_{K_0}^{(0)}(\cdot)
\right)^{\tau}  \tilde{\bb}_i^{(0)},
$$
$$
\tilde{a}_{i,j}(\cdot)
=
1 +
\left(
B_1^{(0)}(\cdot)-\bar{B}_{i,u,1}^{(0)}, \ \cdots, \ B_{K_0}^{(0)}(\cdot) -
\bar{B}_{i,u,K_0}^{(0)}
\right)^{\tau} \tilde{\ba}_{i,j}^{(0)},
$$
$$
\tilde{g}_{i,j}(\cdot)
=
\left(
B_1^{(0)}(\cdot)-\bar{B}_{i,y,j,1}^{(0)}, \ \cdots, \ B_{K_0}^{(0)}(\cdot) -
\bar{B}_{i,y,j,K_0}^{(0)}
\right)^{\tau} \tilde{\bg}_{i,j}^{(0)},
$$
$$
\tilde{a}_{i,p+l}(\cdot)
=
1 + \left(
B_1^{(0)}(\cdot)-\bar{B}_{i,u,1}^{(0)}, \ \cdots, \ B_{K_0}^{(0)}(\cdot) -
\bar{B}_{i,u,K_0}^{(0)}
\right)^{\tau} \tilde{\ba}_{i,p+l}^{(0)},
$$
$$
\tilde{g}_{i,p+l}
=
\left(
B_1^{(0)}(\cdot)-\bar{B}_{i,x,l,1}^{(0)}, \ \cdots, \ B_{K_0}^{(0)}(\cdot) -
\bar{B}_{i,x,l, K_0}^{(0)}
\right)^{\tau}
\tilde{\bg}_{i,p+l}^{(0)},
$$
$i=1, \ \cdots, \ n$, $j=1, \ \cdots, \ p$, $l=1, \ \cdots, \ q$,
as the initial estimators of $b(\cdot) + \eta_i$, $a_{i,j}(\cdot)$,
$g_{i,j}(\cdot)$, $a_{i,p+l}(\cdot)$, and $g_{i,p+l}(\cdot)$, respectively.

\subsubsection{Structure identification}
\label{hom0}
Throughout this paper, for any function $f(\cdot)$ on $[0, 1]$, define $\|f\|_2 = \int_0^1 f^2(v) dv.$ We first estimate the partition $\{\Delta_1, \ \cdots, \ \Delta_m\}$ by using
the idea in \cite{vogt17} and \cite{pei2022network}.  We start with $\tilde{g}_{1,1}(\cdot)$, and compute
\[
\delta_{ij}
=
\frac{1}{\|\tilde{g}_{1,1}\|_2}
\int_0^1 \left\{\tilde{g}_{1,1}(v) - \tilde{g}_{i,j}(v) \right\}^2 dv,
\quad
(i,j) \in \calS,
\]
where $\calS = \{(i,j): i=1, \dots, n; j=1, \dots, p+q\}$.  Let
$\hat{\Delta}_1$ be the set of all $(i,j)$s that satisfy $\delta_{ij} < \calC$, where $\calC$ is a given threshold. In practice, it can be selected by
cross-validation. We select an element, say $(i_0, j_0)$, from
$\calS - \hat{\Delta}_1$, and compute
$$
\delta_{ij}
=
\frac{1}{\|\tilde{g}_{i_0,j_0}\|_2}
\int_0^1 \left\{\tilde{g}_{i_0,j_0}(v) - \tilde{g}_{i,j}(v) \right\}^2 dv,
\quad
(i,j) \in \calS - \hat{\Delta}_1.
$$
Let $\hat{\Delta}_2$ be the set of all $(i,j)$s in $\calS - \hat{\Delta}_1$
that satisfy $\delta_{ij} < \calC$.  Continuously doing so, we get
$\{\hat{\Delta}_1, \dots, \hat{\Delta}_{\hat{m}}\}$ and use it to estimate
$\{\Delta_1, \dots, \Delta_m\}$. Using exactly the same approach, we can get the estimator
$\{\hat{\calD}_1, \dots, \hat{\calD}_{\hat{H}}\}$ of
$\{\calD_1, \dots, \calD_H\}$.

\subsubsection{Final estimation}
\label{fin0}

Let $\bM_i = (\calF_{i,p+1}, \dots, \calF_{i,T_i})^{\tau}$ and $\calF_{i,t}$
be
$$
\bB_{i,0,t}^{\tau} \bb
+
\sum\limits_{j=1}^p
\left(
1 + \bB_{i,u,t}^{\tau} \ba_{i,j}
\right)
\bB_{i,y,j,t}^{\tau} \bg_{i,j}
+
\sum\limits_{l=1}^q
\left(
1 + \bB_{i,u,t}^{\tau} \ba_{i,p+l}
\right)
\bB_{i,x,l,t}^{\tau} \bg_{i,p+l}
$$
with $\ba_{\iota,\ell}$, $\iota=1, \ \cdots, \ n$, $\ell=1, \ \cdots, \ p+q$,
being replaced by $\balpha_k$ if $(\iota,\ell) \in \hat{\calD}_k$,
$\bg_{\iota,\ell}$, $\iota=1, \ \cdots, \ n$, $\ell=1, \ \cdots, \ p+q$, being
replaced by $\bbeta_k$ if $(\iota,\ell) \in \hat{\Delta}_k$.

To take the within cluster correlation into account in the final estimation,
we have to get the estimators of $\sigma_{\eta}^2$ and $\sigma^2$ first.  We
estimate $\sigma_{\eta}^2$ and $\sigma^2$ based on the residuals of working
independence fitting.  Specifically, we first minimise
$$
\sum\limits_{i=1}^n (\bY_i - \bM_i)^{\tau} (\bY_i - \bM_i)
$$
with respect to $(\bb, \ \balpha_1, \dots, \balpha_{\hat{H}}, \ \bbeta_1,
\dots, \bbeta_{\hat{m}})$, and denote the minimiser by
$$
(\tilde{\bb}, \ \tilde{\balpha}_1, \dots, \tilde{\balpha}_{\hat{H}}, \
\tilde{\bbeta}_1, \dots, \tilde{\bbeta}_{\hat{m}}).
$$
Let $\tilde{\bM}_i$ be $\bM_i$ with $\bb$, $\balpha_k$s and $\bbeta_k$s being
replaced by $\tilde{\bb}$, $\tilde{\balpha}_k$s and $\tilde{\bbeta}_k$s.  We then
have the following objective function:
\begin{equation}
\sum\limits_{i=1}^n
\left\|
 (\bY_i - \tilde{\bM}_i) (\bY_i - \tilde{\bM}_i)^{\tau}
-
\sigma^2 I_{T_i} - \sigma_{\eta}^2 \1_{T_i} \1_{T_i}^{\tau}
\right\|_2
\label{sig0}
\end{equation}

Applying the weighted least squares estimation, by simple calculation, we have
the objective function:
\begin{eqnarray}
& & L(\bb, \ \balpha_1, \dots, \balpha_{\hat{H}}, \ \bbeta_1, \ \cdots, \
\bbeta_{\hat{m}})
\nonumber
\\
& = &
\sum\limits_{i=1}^n (\bY_i - \bM_i)^{\tau}
\left(
I_{T_i} -
\frac{\hat{\sigma}_{\eta}^2}{T_i \hat{\sigma}_{\eta}^2 + \hat{\sigma}^2}
\1_{T_i} \1_{T_i}^{\tau}
\right)
(\bY_i - \bM_i)
\label{fin1}
\end{eqnarray}
where $\bY_i = (y_{i,p+1}, \dots, y_{i,T_i})^{\tau}$.

We minimise~(\ref{fin1}) with respect to $(\bb, \balpha_1, \dots, \balpha_{\hat{H}}, \bbeta_1, \dots, \bbeta_{\hat{m}})$, and denote the
minimiser by
$$
(\hat{\bb}, \ \hat{\balpha}_1, \dots, \hat{\balpha}_{\hat{H}}, \
\hat{\bbeta}_1, \dots, \hat{\bbeta}_{\hat{m}}).
$$
The final estimators of $b(\cdot)$, $a_{i,j}(\cdot)$, $g_{i,j}(\cdot)$,
$a_{i,p+l}(\cdot)$ and $g_{i,p+l}(\cdot)$ are
$$
\hat{b}(\cdot)
=
\left(
B_1(\cdot), \dots, B_K(\cdot)
\right)^{\tau}  \hat{\bb},
$$
$$
\hat{a}_{i,j}(\cdot)
=
1 +
\left(
B_1(\cdot)-\bar{B}_{i,u,1}, \dots, B_K(\cdot)-\bar{B}_{i,u,K}
\right)^{\tau} \hat{\balpha}_k, \quad \mbox{if } (i,j) \in \hat{\calD}_k,
$$
$$
\hat{g}_{i,j}(\cdot)
=
\left(
B_1(\cdot)-\bar{B}_{i,y,j,1}, \dots, B_K(\cdot)-\bar{B}_{i,y,j,K}
\right)^{\tau} \hat{\bbeta}_k, \quad \mbox{if } (i,j) \in \hat{\Delta}_k,
$$
$$
\hat{a}_{i,p+l}(\cdot)
=
1 + \left(
B_1(\cdot)-\bar{B}_{i,u,1}, \dots, B_K(\cdot)-\bar{B}_{i,u,K}
\right)^{\tau} \hat{\balpha}_k, \quad \mbox{if } (i,p+l) \in \hat{\calD}_k,
$$
$$
\hat{g}_{i,p+l}
=
\left(
B_1(\cdot)-\bar{B}_{i,x,l,1}, \dots, B_K(\cdot)-\bar{B}_{i,x,l, K}
\right)^{\tau}
\hat{\bbeta}_k, \quad \mbox{if } (i,p+l) \in \hat{\Delta}_k.
$$

\subsection{Computational algorithm}

The main hurdle in the implementation of the proposed estimation method is the
minimisation of (\ref{ls1}) and~(\ref{fin1}). Because neither of them has a
minimiser with closed form, we are going to minimise them by an iterative
approach.

Applying the three-step spline estimation method proposed in \cite{hhy2019}, we
can easily get the initial values for $\bg_{i,j}^{(0)}$'s and
$\bg_{i,p+l}^{(0)}$'s in~(\ref{ls1}).
Replacing $\bg_{i,j}^{(0)}$'s and $\bg_{i,p+l}^{(0)}$'s in~(\ref{ls1}) by their
initial values and minimising~(\ref{ls1}) with respect to $\bb_i^{(0)}$,
$\ba_{i,j}^{(0)}$'s and $\ba_{i,p+l}^{(0)}$'s, we take the resulting minimisers
as the initial values for $\bb_i^{(0)}$, $\ba_{i,j}^{(0)}$'s and
$\ba_{i,p+l}^{(0)}$'s.  Replacing $\bb_i^{(0)}$, $\ba_{i,j}^{(0)}$'s and
$\ba_{i,p+l}^{(0)}$'s in~(\ref{ls1}) by their initial values and minimising
(\ref{ls1}) with respect to $\bg_{i,j}^{(0)}$'s and $\bg_{i,p+l}^{(0)}$'s, we
take the resulting minimisers as the updated values for $\bg_{i,j}^{(0)}$'s and
$\bg_{i,p+l}^{(0)}$'s.  Continuing this iterative process until convergence,
we get the minimiser of~(\ref{ls1}).

The minimisation of~(\ref{fin1}) is similar to that of~(\ref{ls1}).  We start
by choosing the initial values for $\bbeta_k$s in~(\ref{fin1}) based on
$\tilde{\bg}_{i,j}$s and $\tilde{\bg}_{i,p+l}$s obtained in Section~\ref{ini0}.
 Specifically, the initial value for $\bbeta_k$ is taken to be
$$
\frac{1}{|\hat{\Delta}_k|}
\sum\limits_{(\iota,\ell) \in \hat{\Delta}_k} \tilde{\bg}_{\iota,\ell}.
$$
We replace $\bbeta_k$s in~(\ref{fin1}) by their initial values, then minimise~(\ref{fin1}) with respect to $\bb$ and $\balpha_k$s.  We take the resulting
minimisers as the initial values for $\bb$ and $\balpha_k$s and substitute
them for $\bb$ and $\balpha_k$s in~(\ref{fin1}), then minimise~(\ref{fin1})
with respect to $\bbeta_k$'s.  We take the resulting minimisers as the updated
values for $\bbeta_k$s, and continue this iterative process until convergence
to get the minimiser of~(\ref{fin1}).

\subsection{A couple of remarks}

\begin{remark}\label{rem1}
The estimation of the partition in Section~\ref{hom0} can be
further improved.  Taking $\{\hat{\Delta}_1, \dots,
\hat{\Delta}_{\hat{m}}\}$ as an example, we can apply the following iterative
process to improve the estimation:
\begin{enumerate}
\item[(1)] Compute
$$
\bar{g}_k(\cdot)
=
\frac{1}{|\hat{\Delta}_k|}
\sum\limits_{(i,j) \in \hat{\Delta}_k} \tilde{g}_{i,j}(\cdot),
\quad
k = 1, \dots, \hat{m},
$$
where $|\hat{\Delta}_k|$ is the cardinality of $\hat{\Delta}_k$.

\item[(2)] For each $(i,j) \in \calS$, we compute
$$
c_k = \int_0^1 \left\{\bar{g}_k(v) - \tilde{g}_{i,j}(v) \right\}^2 dv,
\quad
k = 1, \dots, \hat{m}.
$$
If $c_{k_0}$ is the smallest $c_k$, then $(i,j)$ belongs to set
$\hat{\Delta}_{k_0}^{(1)}$.  This leads to a partition
$\{\hat{\Delta}_1^{(1)}, \dots, \hat{\Delta}_{\hat{m}}^{(1)}\}$ of
$\calS$.

\item[(3)] We treat $\{\hat{\Delta}_1^{(1)}, \dots,
\hat{\Delta}_{\hat{m}}^{(1)}\}$ as $\{\hat{\Delta}_1, \dots,
\hat{\Delta}_{\hat{m}}\}$, and repeat (1) and (2).  Continuously doing so until
convergence, we get an improved estimator of the partition $\{\Delta_1, \
\cdots, \ \Delta_m\}$.
\end{enumerate}
\end{remark}

\begin{remark}\label{rem2}
In the final estimation, an iterative process can be used
to further improve the final estimators.  Specifically, we substitute the
minimiser of (\ref{fin1}) for $\tilde{\bb}$, $\tilde{\balpha}_k$s and
$\tilde{\bbeta}_k$s in (\ref{sig0}), and minimise (\ref{sig0}).  We treat the
resulting minimiser as updated $\hat{\sigma}_{\eta}^2$ and $\hat{\sigma}^2$,
and substitute them for the $\hat{\sigma}_{\eta}^2$ and $\hat{\sigma}^2$ in
(\ref{fin1}), then minimise (\ref{fin1}) and substitute the resulting minimiser
for $\tilde{\bb}$, $\tilde{\balpha}_k$s and $\tilde{\bbeta}_k$s in
(\ref{sig0}), and minimise (\ref{sig0}) to get updated $\hat{\sigma}_{\eta}^2$
and $\hat{\sigma}^2$.  Continue this iterative process until convergence, and
substitute the converged $\hat{\sigma}_{\eta}^2$ and $\hat{\sigma}^2$ for the
$\hat{\sigma}_{\eta}^2$ and $\hat{\sigma}^2$ in (\ref{fin1}), then minimise
(\ref{fin1}) to get the final $\hat{\bb}$, $\hat{\balpha}_k$s and
$\hat{\bbeta}_k$s based on which improved final estimators of $b(\cdot)$,
$a_{i,j}(\cdot)$, $g_{i,j}(\cdot)$, $a_{i,p+l}(\cdot)$, and $g_{i,p+l}(\cdot)$
are obtained.
\end{remark}

\subsection{Selecting the values of the tuning parameters}

In this section, we address several practical problems regarding the selection of the values of the tuning parameters in the proposed methods.

\subsubsection{Selecting the optimal number of knots in the spline decomposition}

Following the lead of \cite{hhy2019}, we select the number of knots $K_{0}$ via a Bayesian information criterion (BIC) approach:
\[
\mathrm{BIC}(K_{0})=\log \left(\mathrm{RSS}\right) + \mathcal{N}\frac{\log T}{T}.
\]
The residual sum of squares ($\mathrm{RSS}$) is defined in~(\ref{ls1}), which reflects the goodness of fit. Here, $\mathcal{N}= (2p+2q+1)K_{0}$ controls the complexity of the model, and we assume $T_{i}=T$ for the sake of clarification. The optimal number of knots $K_{0}$ can be estimated by minimising the above BIC. According to Theorem~\ref{thm:1}, it is reasonable to choose the optimal number of knots $K_{0}$ on the interval $\lfloor 0.5T^{1/5}, 2T^{1/5}\rfloor$, where $\lfloor a \rfloor$ denotes the largest integer not larger than $a$. The number of knots $K$ in the final estimation can be selected using the same approach.

\subsubsection{Selecting the threshold $\calC$ in the structure identification}

The structure identification method relies on prior information about the threshold $\calC$. However, it is usually unknown in practical applications and needs to be determined via a data-driven method.  Actually, selecting $\calC$ is equivalent to selecting $\hat{H}$ and $\hat{m}$. In this paper, following the lead of \cite{lqz2021}, we can also apply the cross-validation method to select the two tuning parameters, $\hat{H}$ and $\hat{m}$.

In particular, we implement a $V$-fold cross-validation approach. For a given pair $\left\{H, m\right\}$, we remove the $1 / V$th member of the observed time points for
\[
\left\{\left(y_{i, t}, u_{i,t},\mathbf{X}_{i, t}\right), i=1, \dots, n, t= 1, \dots, T_{i}\right\}
\]
as a validation set. We estimate the semiparametric model~(\ref{model}) with identified structure on the remaining data, compute the squared error between $y_{i,t}$ and the fitted values on the validation set, and repeat this procedure $V$ times to calculate the cross-validated mean squared error. The optimal $\hat{H}$ and $\hat{m}$ can be estimated by minimising the cross-validation mean squared error.

\section{Asymptotic theory}
\label{asy0}

In this section, we are going to demonstrate the advantage of the proposed
methodology by asymptotic theory.  We will consider three different approaches:

\begin{itemize}
\item {\it Over-fitting}:  Treat all unknown functions in model (\ref{model})
as different functions to estimate.  Namely, directly apply the final
estimation described in Section \ref{fin0} under the assumption that either
of the two partitions involved is $\{\{(i,j)\}: \ i=1, \ \cdots, \ n; \ j=1, \
\cdots, \ p+q\}$.

\item {\it Under-fitting}: Directly apply the final estimation to estimate the
unknown functions under the assumption
$$
a_{1,j}(\cdot) = \cdots = a_{n,j}(\cdot),
\quad
g_{1,j}(\cdot) = \cdots = g_{n,j}(\cdot),
$$
for $j=1, \dots, p+q$.

\item {\it Correct fitting}\/: The proposed estimation is used to estimate the
unknown functions.
\end{itemize}

In this section, we assume $T_{i} \rightarrow \infty$, $n$ possibly diverges to
infinity, but $H$ and $m$ are fixed.  This agrees with many real applications
where $H$ and $m$ are expected to be small, and thus, there is a significant
reduction of the unknown functions by clustering similar functions.

To keep the presentation concise, we state the asymptotic theorems in this
section and leave all technical proofs in the supplementary material. Let
$$
\bB^{(0)}(u)
=
\left(
B_1^{(0)}(u), \dots, B_{K_0}^{(0)}(u)
\right)^{\tau},
$$
$$
\bB_{i}^{(0)}(u)
=
\left(
B_1^{(0)}(u)-\bar{B}_{i,u,1}^{(0)}, \dots, B_{K_0}^{(0)}(u) -
\bar{B}_{i,u,K_0}^{(0)}
\right)^{\tau},
$$
$$
\bB^{(0)}_{i,j}(x)
=
\left(
B_1^{(0)}(x)-\bar{B}_{i,x,j,1}^{(0)}, \dots, B_{K_0}^{(0)}(x) -
\bar{B}_{i,x,j,K_0}^{(0)}
\right)^{\tau},
$$
$$
\mathbf{\tilde{A}}_{i,j}(u)
=
(\mathbf{0}^{\tau}, \dots, \bB_{i}^{(0)\tau}(u), \dots,
\mathbf{0}^{\tau})^{\tau},
\quad
\mathbf{\tilde{G}}_{i,j}(x)
=
(\mathbf{0}^{\tau}, \dots, \bB^{(0)\tau}_{i,j}(x), \dots,
\mathbf{0}^{\tau})^{\tau}.
$$
$\mathbf{\tilde{A}}_{i,j}(u)$ is a $\{(p+q+1)K_{0}\}$-dimensional vector
consisting of $(p+q+1)$ blocks of length $K_0$, with the $(j+1)$th block being
$\bB_{i}^{(0)\tau}(u)$, others all being $\0$.  $\mathbf{\tilde{G}}_{i,j}(x)$
is a $\{(p+q)K_{0}\}$-dimensional vector consisting of $(p+q)$ blocks of
length $K_0$, with the $j$th block being $\bB^{(0)\tau}_{i,j}(x)$, others all
being $\0$.

\begin{theorem}[Over-fitting case]\label{thm:1}
For any $i$, $i=1,\dots,n$, and $1\leq j \leq p+q$, under the technical
conditions (C1)--(C3) and (C5) in the supplementary material, if $K_{0}=O(T_{i}^{1/5})$, we have
$$
T_i^{2/5} \left(\hat{b}(u) -  b(u) - r(u) \right)
\stackrel{D}{\longrightarrow}
N\left(
0, \ \mathbf{\tilde{A}}^{\tau}_{i,0}(u) \tilde{\Xi}_{1}^{-1}
\tilde{\Sigma}_{1} \tilde{\Xi}_{1}^{-1} \mathbf{\tilde{A}}_{i,0}(u)
\right),
$$
$$
T_i^{2/5} \Big(\hat{a}_{i,j}(u) -  a_{i,j}(u)-r_{i,j}(u) \Big) \stackrel{D}{\longrightarrow} N(0, \mathbf{\tilde{A}}^{\tau}_{i,j}(u) \tilde{\Xi}_{1}^{-1} \tilde{\Sigma}_{1} \tilde{\Xi}_{1}^{-1} \mathbf{\tilde{A}}_{i,j}(u)),
$$
$$
T_i^{2/5} \Big(\hat{g}_{i,j}(x) -  g_{i,j}(x) - d_{i,j}(x) \Big) \stackrel{D}{\longrightarrow} N(0, \mathbf{\tilde{G}}^{\tau}_{i,j}\left(x\right) \tilde{\Xi}_{2}^{-1} \tilde{\Sigma}_{2} \tilde{\Xi}_{2}^{-1} \mathbf{\tilde{G}}_{i,j}\left(x\right)),
$$
where $\hat{b}(u)$, $\hat{a}_{i,j}(u)$ and $\hat{g}_{i,j}(x)$ are obtained by overfitting, the bias terms $r(u)=b(u)-\bB^{(0)\tau}(u)\bb_i^{(0)}$, $r_{i,j}(u)=a_{i,j}(u)-\bB_{i}^{(0)\tau}(u)\ba_{i,j}^{(0)}$,  $d_{i,j}(x)=g_{i,j}(x)-\bB_{i,j}^{(0)\tau}(x)\bg_{i,j}^{(0)}$ are of order $O(K^{-2}_{0})$.  $\tilde{\Xi}_{1}$ and $\tilde{\Xi}_{2}$ are defined in condition (C5). $\tilde{\Sigma}_{1}=(\sigma^2+\sigma^2_{\eta})\int_{0}^{1} \mathrm{E}\left[\tilde{\bD}_{i}(u) \tilde{\bD}^{\tau}_{i}(u) \right] \mathrm{d} u $, $\tilde{\Sigma}_{2}=(\sigma^2+\sigma^2_{\eta})\int_{0}^{1} \mathrm{E}\left[\tilde{\bZ}_{i}(x) \tilde{\bZ}^{\tau}_{i}(x) \right] \mathrm{d} x$, and
$$
\tilde{\bD}_{i}(u)=\left(\bB^{(0)\tau}(u), g_{i,1}(x_{i,t,1}) \bB_{i}^{(0)\tau}(u), \dots, g_{i,p+q}(x_{i,t,p+q}) \bB_{i}^{(0)\tau}(u)\right)^{\tau},
$$
$$
\tilde{\bZ}_{i}(x) = \left(a_{i,1}(u_{i,t}) \bB_{i,1}^{(0)\tau}(x), \dots, a_{i,p+q}(u_{i,t}) \bB_{i,p+q}^{(0)\tau}(x)\right)^{\tau}.
$$
\end{theorem}

From Theorem~\ref{thm:1}, it is easy to see that the convergence rate of the
estimators $\hat{b}(u)$, $\hat{a}_{i,j}(u)$ and $\hat{g}_{i,j}(x)$ are of
order $T_{i}^{-2 / 5}$, which is as expected, as we assume that the functions
are twice differentiable.

\begin{theorem}[Under-fitting case]\label{thm:2}
Suppose the functions $a_{i,j}(\cdot)$s are sufficiently separated, i.e.,
$$
\frac{1}{n}\sum\limits_{i=1}^n \int_0^1 \{a_{i,j}(u) - \bar{a}_1(u)\}^2 du
> C > 0,
\quad
\bar{a}_1(u) = \frac{1}{n}\sum\limits_{i=1}^n a_{i,j}(u),
$$
then we have
$$
\|\hat{a}_{i,j} -  a_{i,j}\|_2 > C.
$$
Similarly, we have
$$
\|\hat{b} -  b\|_2 > C,
\quad
\|\hat{g}_{i,j} -  g_{i,j}\|_2 > C,
$$
where $\hat{b}(u)$, $\hat{a}_{i,j}(u)$ and $\hat{g}_{i,j}(x)$ are obtained by
underfitting.
\end{theorem}

Theorem~\ref{thm:2} shows that the estimators obtained by underfitting are not
consistent. Before presenting Theorem~\ref{thm:3}, we introduce some notations: let
$$
\bB(u)
=
\left(
B_1(u), \dots, B_{K}(u)
\right)^{\tau},
\quad
\bB_{i}(u)
=
\left(
B_1(u)-\bar{B}_{i,u,1}, \dots, B_{K}(u) -
\bar{B}_{i,u,K}
\right)^{\tau},
$$
$$
\bB_{i,j}(x)
=
\left(
B_1(x)-\bar{B}_{i,x,j,1}, \dots, B_{K}(x) -
\bar{B}_{i,x,j,K}
\right)^{\tau},
$$
$$
\mathbf{A}_{i,j}(u)
=
(\mathbf{0}^{\tau}, \dots, \bB_{i}^{\tau}(u), \dots, \mathbf{0}^{\tau})^{\tau},
\quad
\mathbf{G}_{i,j}(x) = (\mathbf{0}^{\tau}, \dots, \bB^{\tau}_{i,j}(x), \dots, \mathbf{0}^{\tau})^{\tau}.
$$
$\mathbf{A}_{i,j}(u)$ is a $\{(p+q+1)K\}$-vector consisting of $(p+q+1)$ blocks
of length $K$, with the $(j+1)$th block being $\bB_{i}^{\tau}(u)$, others all
being $\0$.  $\mathbf{G}_{i,j}(x)$ is a $\{(p+q)K\}$-vector consisting of
$(p+q)$ blocks of length $K$, with the $j$th block being $\bB^{\tau}_{i,j}(x)$,
others all being $\0$.

\begin{theorem}[Correct fitting case]\label{thm:3}
 Under the technical conditions
(C1)--(C5) in the supplementary material, let $N = \sum_{i=1}^n T_i$. When
$K_0 = O(T_i^{1/5})$ and $K = O(N^{1/5})$, we have
$$
N^{2/5} \left(\hat{b}(u) -  b(u)- r(u) \right)
\stackrel{D}{\longrightarrow}
N\left(
0, \ \mathbf{A}^{\tau}_{i,0}(u) \Xi_{1}^{-1} \Sigma_{1} \Xi_{1}^{-1} \mathbf{A}_{i,0}(u)
\right),
$$
$$
N^{2/5} \Big(\hat{a}_{i,j}(u) -  a_{i,j}(u) - r_{i,j}(u) \Big)
\stackrel{D}{\longrightarrow}
N\left(0, \ \mathbf{A}^{\tau}_{i,j}(u) \Xi_{1}^{-1} \Sigma_{1} \Xi_{1}^{-1} \mathbf{A}_{i,j}(u)
\right),
$$
$$
N^{2/5} \Big(\hat{g}_{i,j}(x) -  g_{i,j}(x) - d_{i,j}(x) \Big)
\stackrel{D}{\longrightarrow}
N\left(0, \ \mathbf{G}^{\tau}_{i,j}\left(x\right) \Xi_{2}^{-1} \Sigma_{2} \Xi_{2}^{-1} \mathbf{G}_{i,j}\left(x\right)
\right),
$$
where $\hat{b}(u)$, $\hat{a}_{i,j}(u)$ and $\hat{g}_{i,j}(x)$ are obtained by
correct fitting, and the bias terms $r(u), r_{i,j}(u), d_{i,j}(x)$ are defined as Theorem 1 with order of $K^{-2}$. $\Xi_{1}$ and $\Xi_{2}$ are defined in condition (C5), and
$$
\Sigma_{1}
=
(\sigma^2+\sigma^2_{\eta})\int_{0}^{1} \mathrm{E}
\left[\bD_{k}(u) \bD_{k}(u)^{\tau} \right] \mathrm{d} u,
\quad
\Sigma_{2}
=
(\sigma^2+\sigma^2_{\eta})\int_{0}^{1} \mathrm{E}
\left[\bZ_{k}(x) \bZ_{k}(x)^{\tau} \right] \mathrm{d} x,
$$
$$
\bD_{k}(u)=\left(\bB^{\tau}(u), \beta_{k}(x_{i,t,1}) \bB_{i}^{\tau}(u), \dots,\beta_{k}(x_{i,t,p+q}) \bB_{i}^{\tau}(u)\right)^{\tau}, \mbox{if } (i,j) \in \calD_k,
$$
$$
\bZ_{k}(x) = \left(\alpha_{k}(u_{i,t}) \bB_{i,1}^{\tau}(x), \dots,\alpha_{k}(u_{i,t}) \bB_{i,p+q}^{\tau}(x)\right)^{\tau},\mbox{if } (i,j) \in \Delta_k.
$$
\end{theorem}

Theorem~\ref{thm:3} shows that the convergence rates of the estimators $\hat{b}(u)$, $\hat{a}_{i,j}(u)$ and $\hat{g}_{i,j}(x)$ are of order $N^{-2/5}$. This together with Theorem~\ref{thm:1} shows that the estimators obtained by correct fitting have convergence rates with a higher order than those obtained by overfitting. Therefore, they are more accurate.

Because the asymptotic variance of the above estimators has a very complicated form and it is not clear how to estimate it consistently, constructing a statistical inference based on the asymptotic normality established in Theorem~\ref{thm:3} can be very challenging. In this paper, we do not consider the inference problem and leave it as an open question.

\section{Simulations}
\label{simu0}

In this section, we use a simulated example to demonstrate how well the proposed estimation procedure works and the risk of ignoring the homogeneity or heterogeneity among individuals.

In particular, the data are generated from the following semiparametric clustered data model:
$$
y_{i,t}
=
b(u_{i,t})
+
 a_{i,1}(u_{i,t}) g_{i,1}(y_{i,t-1})
+
a_{i,2}(u_{i,t}) g_{i,2}(x_{i,t})
 +
\eta_i
+
\epsilon_{i,t},
$$
where $u_{i,t}$ are generated from a uniform distribution $U(0,1)$. The exogenous covariates are generated via a first-order autoregressive (AR) process with parameters $\rho$ and $\sigma$, denoted by $\operatorname{AR}(1 ; \rho, \sigma)$, which is given by $x_{i,t}=\rho  x_{i,t-1}+\sigma \zeta_{i,t}$, where $\zeta_{i,t}$ are generated from i.i.d.\ standard normal random variables. In particular, $ x_{i,t} \sim \operatorname{AR}(1 ; 0.6,0.5)$, and $\epsilon_{i,t}$  and $\eta_i$ are independently generated from $N(0,0.1^2)$.

\begin{figure}
\centering
\begin{subfigure}{.45\textwidth}
  \centering
  \includegraphics[width=0.8\textwidth]{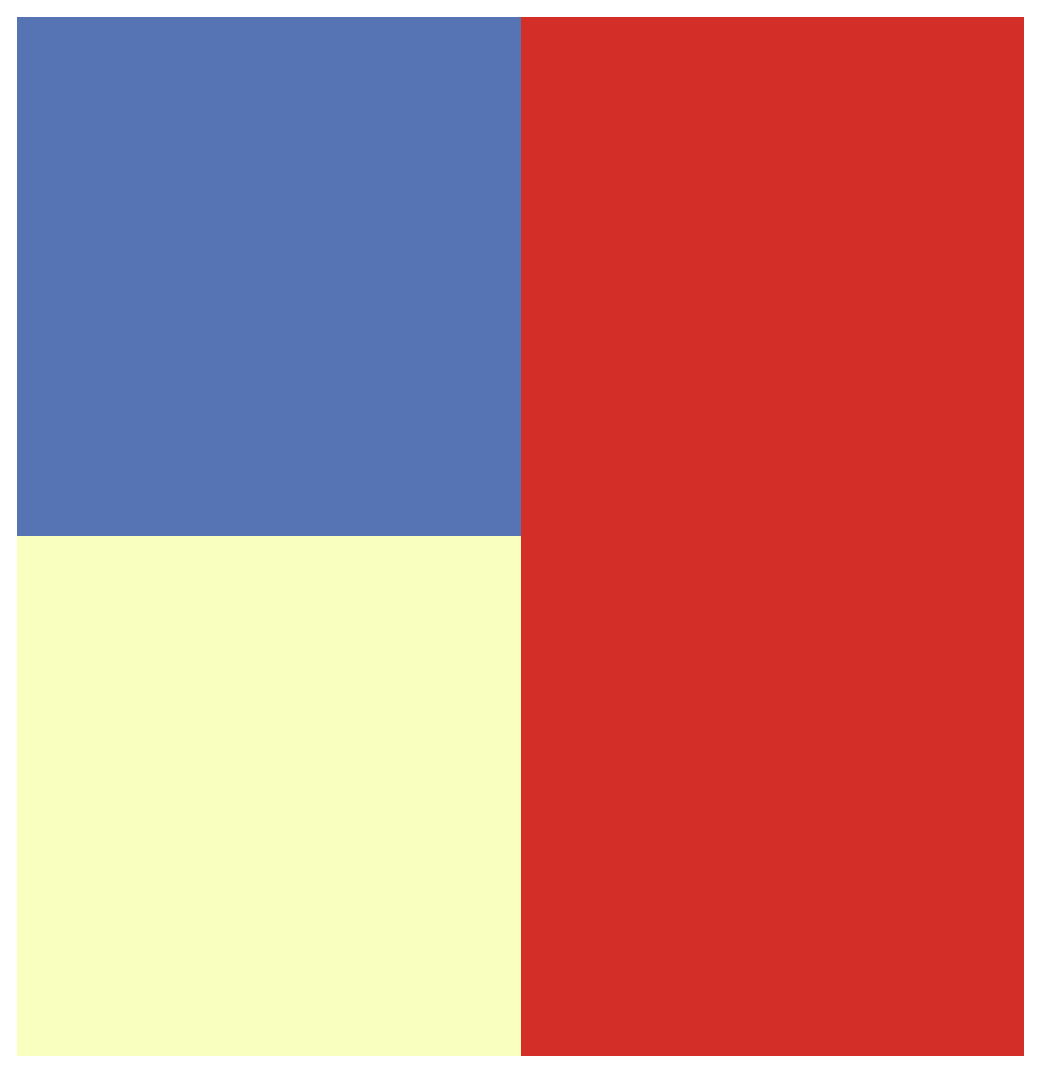}
  \caption{}
\end{subfigure}%
\begin{subfigure}{.45\textwidth}
  \centering
  \includegraphics[width=0.8\textwidth]{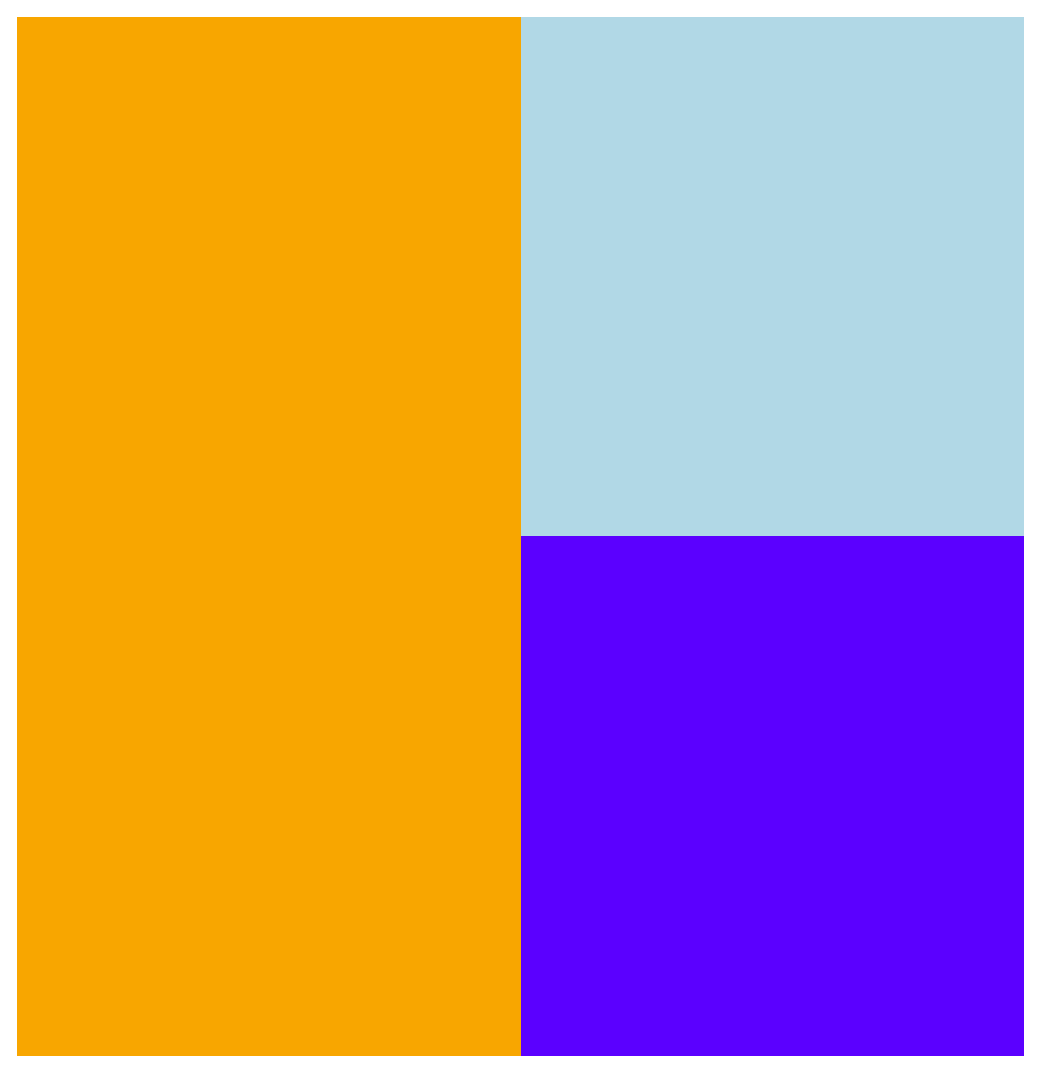}
  \caption{}
\end{subfigure}
\caption{Toy example of the latent structures imposed on (a) $a_{i,j}(\cdot)$ and (b) $g_{i,j}(\cdot)$.}
\label{fig:toy}
\end{figure}

The cluster-specific coefficient functions and additive functions are generated as follows:
$b(u)=1.5 \cos (2 \pi u)$, and for each $i = 1,2, \dots, n $,
\[
a_{i,1}(u)=
\left\{\begin{array}{ll}
1.3 u \sin (2 \pi u)+1, & \text { when } i=1,2, \cdots, n / 2,\\
1.3 u \cos (2 \pi u)+1, & \text { when } i=n / 2+1, \cdots, n.
\end{array}\right.
\]
\[
a_{i,2}(u) = 2 \sin (1.5 \pi u)-1.2(u-0.5)(1-u)+1,
\]
and
\[
g_{i,1}\left(y_{i,t-1}\right) = -0.8(1 - y^2_{i,t-1})/(1 + y^2_{i,t-1}),
\]
\[
g_{i,2}\left(x_{i,t}\right)=
\left\{\begin{array}{ll}
2 \cos \left(\pi x_{i,t} / 2\right)+1.8 \sin \left(\pi x_{i,t} / 3\right), & \text { when } i=1,2, \cdots, n / 2,\\
1.5 \sin \left(\pi x_{i,t} / 4\right)-1.2 \cos \left(\pi x_{i,t} / 3\right), & \text { when } i = n/2+1, \cdots, n.
\end{array}\right.
\]
For the latent structure,  Figure~\ref{fig:toy} describes a toy example of $a_{i,j}(\cdot)$ and $g_{i,j}(\cdot)$ ($i=1, \dots,n$ and $j=1,2$), where different colours denote different functions.

We run the simulated example 100 times with various $n$ and $T$ and compare our proposed approach to its potential competitors based on the following performance metrics:

\emph{Estimation accuracy}.  For  an estimator $\hat{a}_{i,j}(\cdot)$ of $a_{i,j}(\cdot)$, its estimation accuracy can be evaluated based on the mean integrated squared error:
\[
\operatorname{MISE}\left(\hat{a}_{i,j}\right)=E\left\{\int\left(\hat{a}_{i,j}(u)- a_{i,j}(u)\right)^{2} d u\right\}.
\]
To prevent the performance from being dominated by the poor boundary behaviour, we let the integral domain be a non-boundary region, which is between the 1st and 99th quantiles of $\left\{u_{i t}\right\}$.

\emph{Consistency of the structure identification}. To evaluate the distance between
the detected structure and the true one, we use the normalised
mutual information (NMI) \citep{klz2016}, which measures the similarity between
two partitions. Suppose that $\mathbb{C}=\left\{C_{1}, C_{2}, \dots\right\}$
and $\mathbb{D}=\left\{D_{1}, D_{2}, \dots\right\}$ are two partitions of
$\{1, \dots, n\}$, the $\mathrm{NMI}$ is defined as
$$
\mathrm{NMI}(\mathbb{C}, \mathbb{D})=\frac{I(\mathbb{C}, D)}{[H(\mathbb{C})+H(\mathbb{D})] / 2},
$$
where
\[
I(\mathbb{C}, \mathbb{D})=\sum_{k, j}\left(\left|C_{k} \cap D_{j}\right| / n\right) \log \left(n\left|C_{k} \cap D_{j}\right| /\left|C_{k}\right|\left|D_{j}\right|\right),
\]
and
$$
H(\mathbb{C})=-\sum_{k}\left(\left|C_{k}\right| / n\right) \log \left(\left|C_{k}\right| / n\right).
$$
The NMI takes values in $[0, 1]$ with larger values indicating a higher level of similarity between the two partitions. For an estimated partition $\hat{\calD}=\left\{\calD_{1}, \dots, \calD_{\hat{H}}\right\}$ of $\{(i, j): 1, \dots, n, j=1, \dots,p+q\}$, obtained in stage 2 of the proposed estimation procedure in Section~\ref{met0}, we calculate $\mathrm{NMI}(\hat{\calD}, \calD)$ to assess how close the true structure in $a_{i,j}(\cdot)$ is to the estimated one. Similarly, we can calculate $\mathrm{NMI}(\hat{\Delta}, \Delta)$ to assess how close the true structure in $g_{i,j}(\cdot)$ is to the estimated one.

The sample size $T_i = 100$, $200$ or $400$ and $n = 20$ or $40$.  We estimate
the unknown functions by overfitting, correct fitting and underfitting,
respectively.  The mean and standard deviation of NMI for the resulting
estimators are presented in Table~\ref{Table:nmi} and the mean squared errors
of the resulting estimators are presented in Table~\ref{Table:mse}.

\subsection{Initial estimation}

\begin{figure}
\centering
\includegraphics[width=1.0\textwidth,height=8cm]{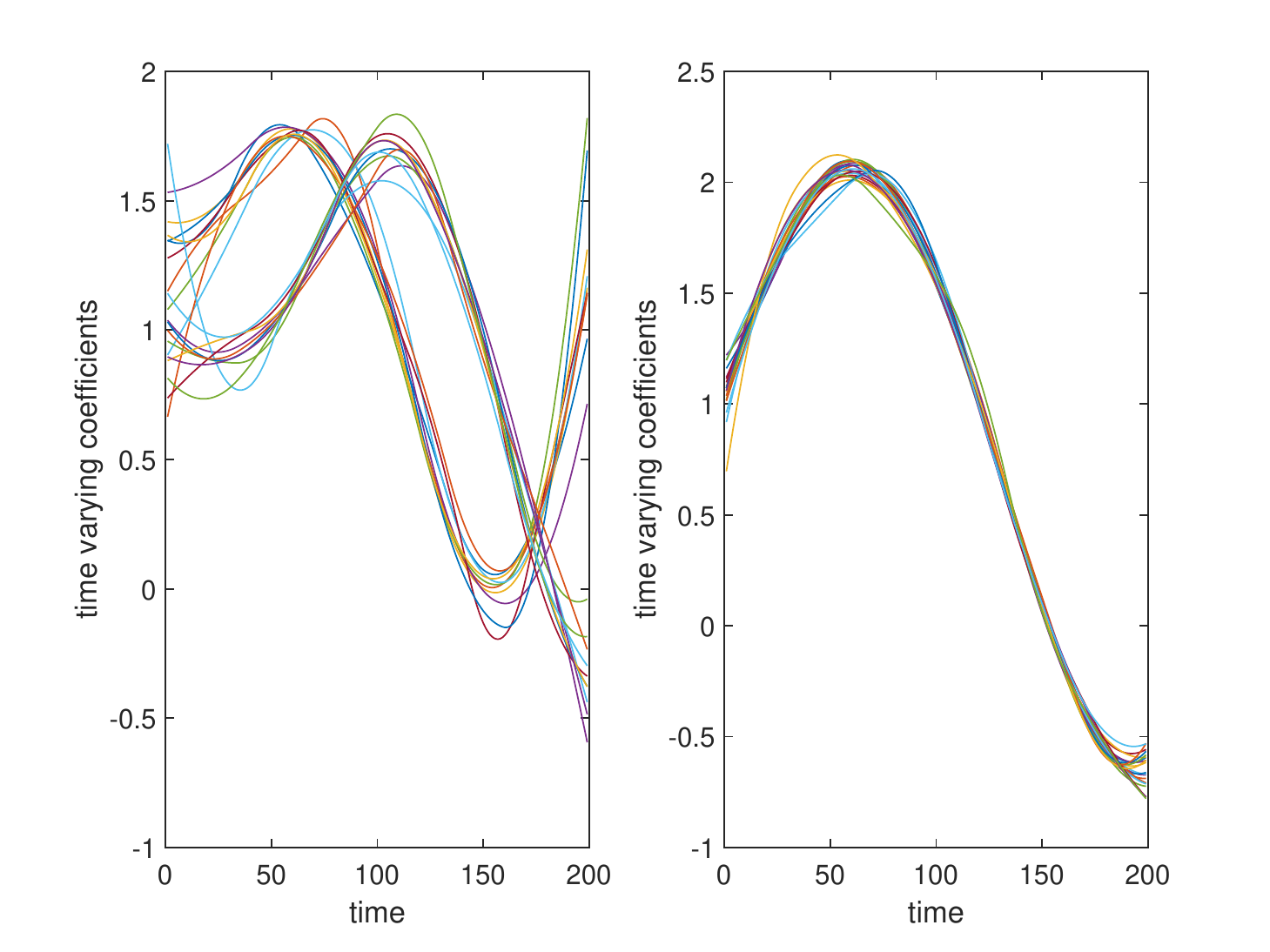}
\caption{Median estimated varying-coefficient functions $a_{i,j}(\cdot)$ based on 100 simulations when $n=20$ and $T=200$. Left: Estimates of $a_{i,1}(\cdot)$. Right: Estimates of $a_{i,2}(\cdot)$.}
\label{fig:initial_alpha}
\end{figure}

\begin{figure}
\centering
\includegraphics[width=1.0\textwidth,height=8cm]{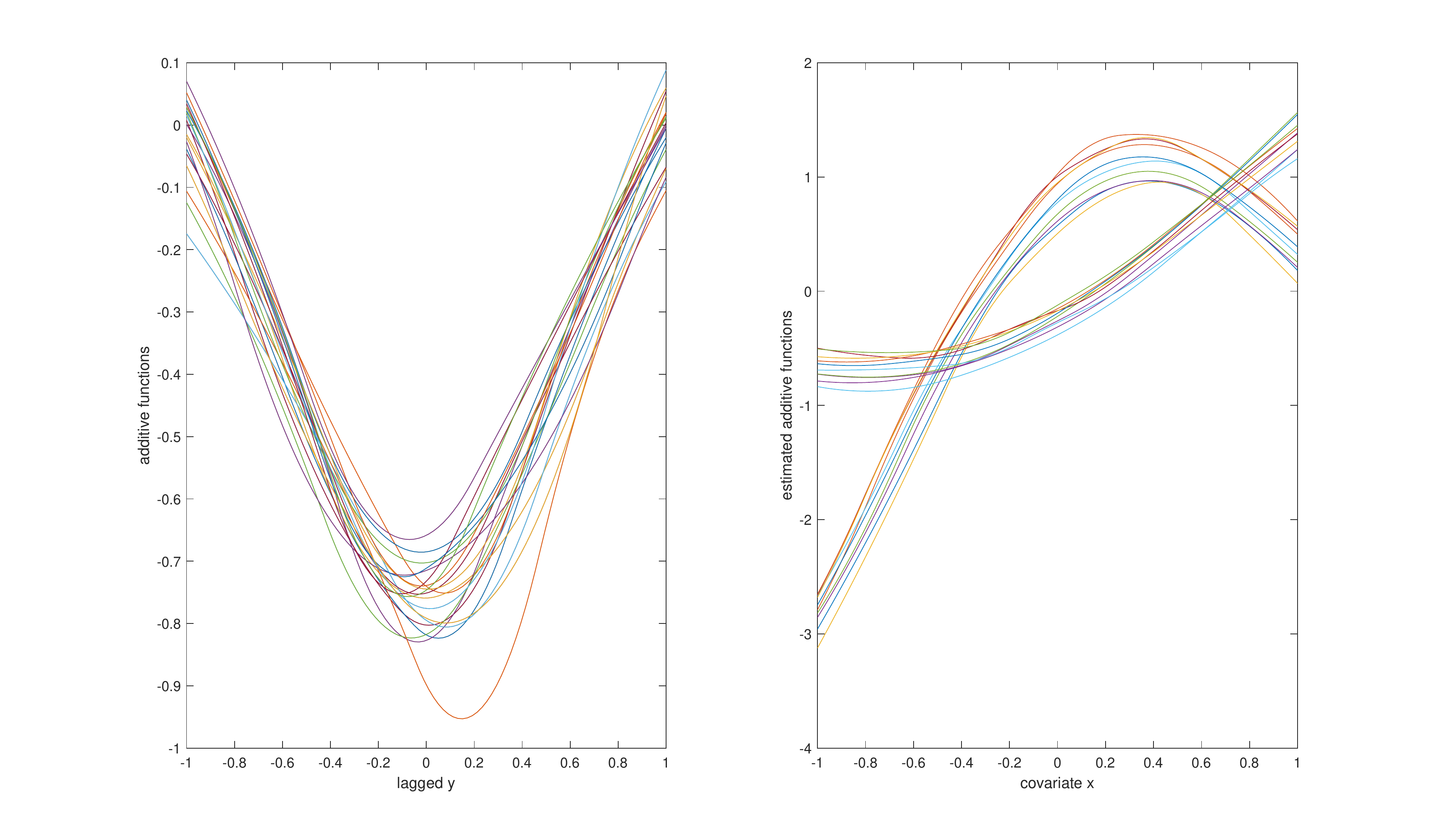}
\caption{Median estimated additive functions $g_{i,j}(\cdot)$ based on 100 simulations when $n=20$ and $T=200$. Left: Estimates of $g_{i,1}(\cdot)$. Right: Estimates of $g_{i,2}(\cdot)$.}
\label{fig:initial_beta}
\end{figure}

The initial estimates for subject-specific varying-coefficient functions $a_{i,j}(\cdot)$ and additive functions $g_{i,j}(\cdot)$ are presented in Figures~\ref{fig:initial_alpha} and~\ref{fig:initial_beta}. Note that the plotted median estimated functions correspond to the simulation which has the median MISEs. From these figures, we can see that the initial estimates for $a_{i,1}(\cdot)$ and $g_{i,2}(\cdot)$ have an obvious group structure, which is a preliminary validation of our simulation setup.

\subsection{Structure identification}

\begin{figure}[!ht]
\centering
\begin{subfigure}{.33\textwidth}
  \centering
  \includegraphics[width=1.0\textwidth]{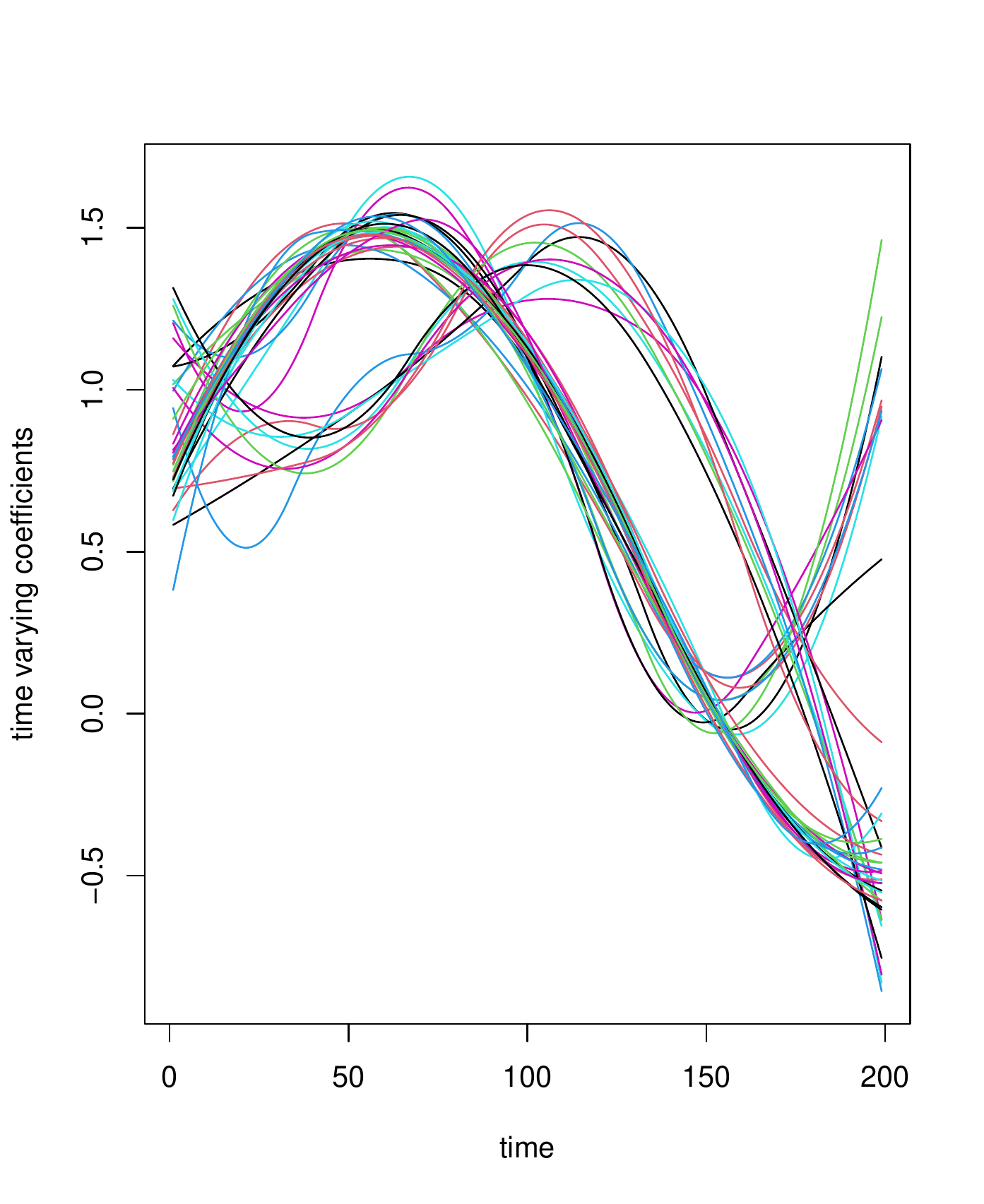}
  \caption{Coefficient functions}
\end{subfigure}%
\begin{subfigure}{.33\textwidth}
  \centering
  \includegraphics[width=1.0\textwidth]{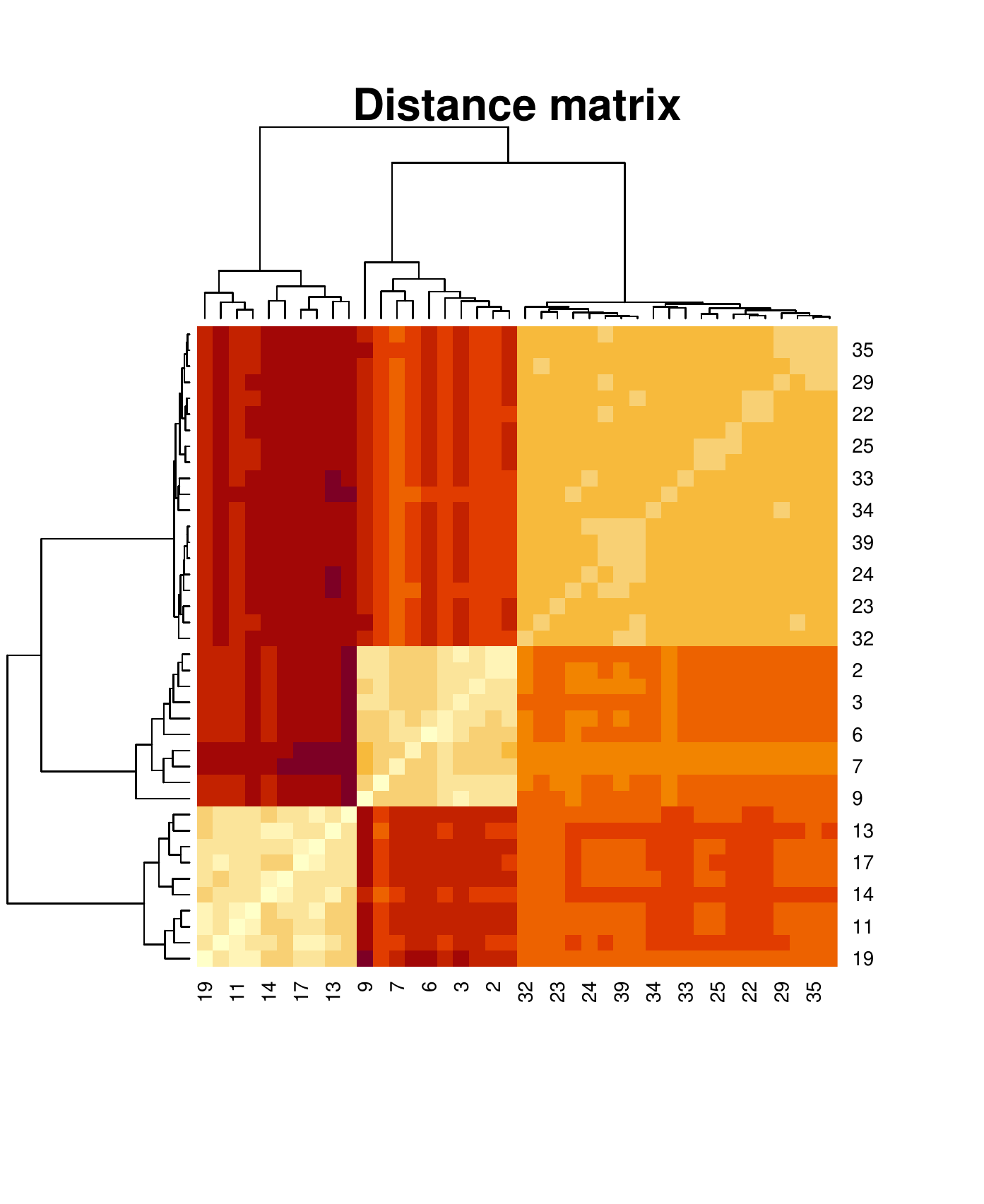}
  \caption{Distance matrix}
\end{subfigure}
\begin{subfigure}{.33\textwidth}
  \centering
  \includegraphics[width=1.0\textwidth]{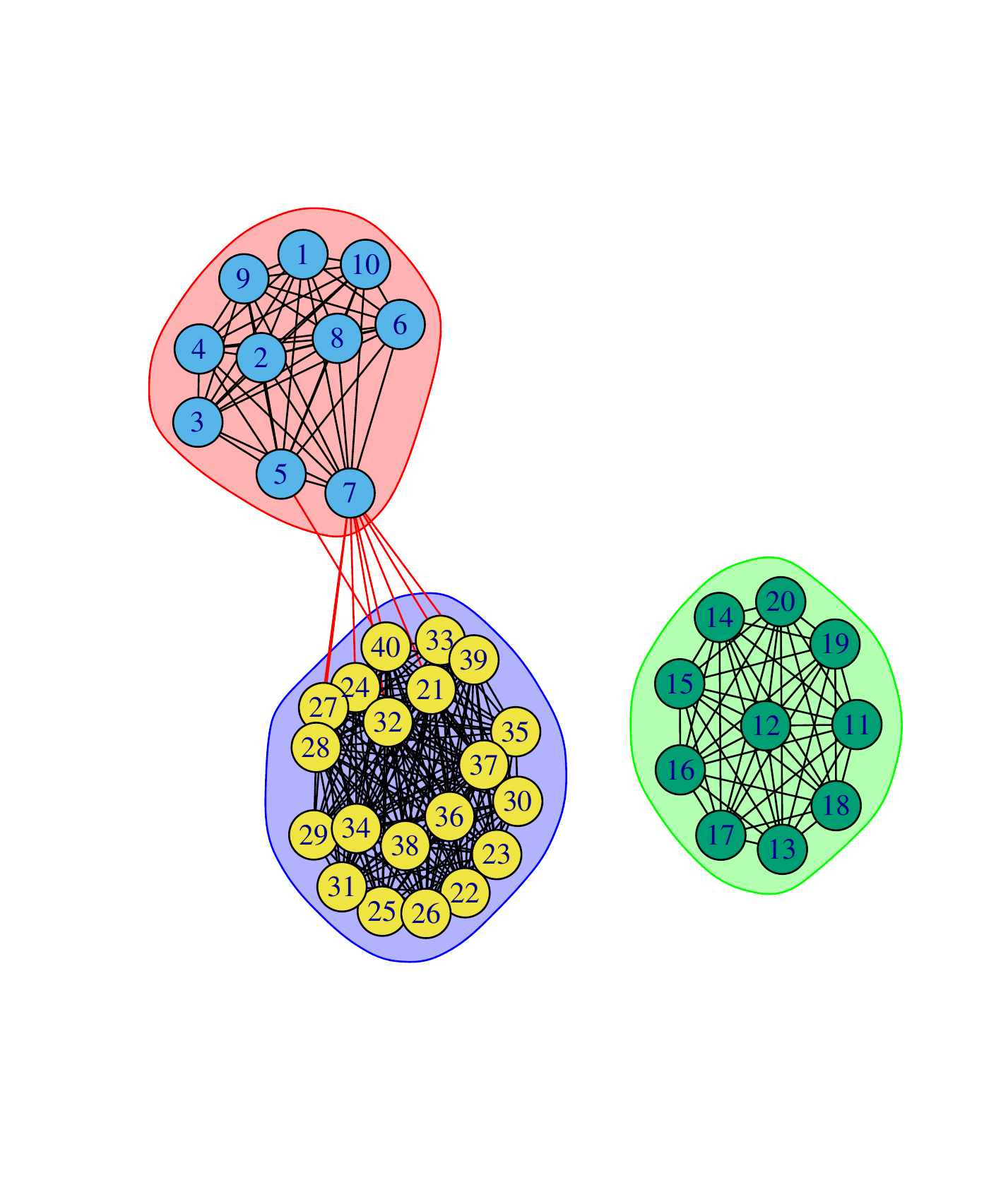}
  \caption{Group structure}
\end{subfigure}
\caption{One typical run of structure identification for $a_{i,j}(\cdot)$, where $i=1,\ldots,20; j=1,2$. (a) The estimates, $\hat{a}_{i,j}(\cdot)$, of specific coefficient functions for each subject $i$. (b) The distance matrix constructed by the $L_{2}$ distance. (c) The identified structure, where the numbers ($1,\ldots,40$) denote the combinations of $(i,j)$.}
\label{fig:flow1}
\end{figure}

\begin{figure}[!ht]
\centering
\begin{subfigure}{.33\textwidth}
  \centering
  \includegraphics[width=1.0\textwidth]{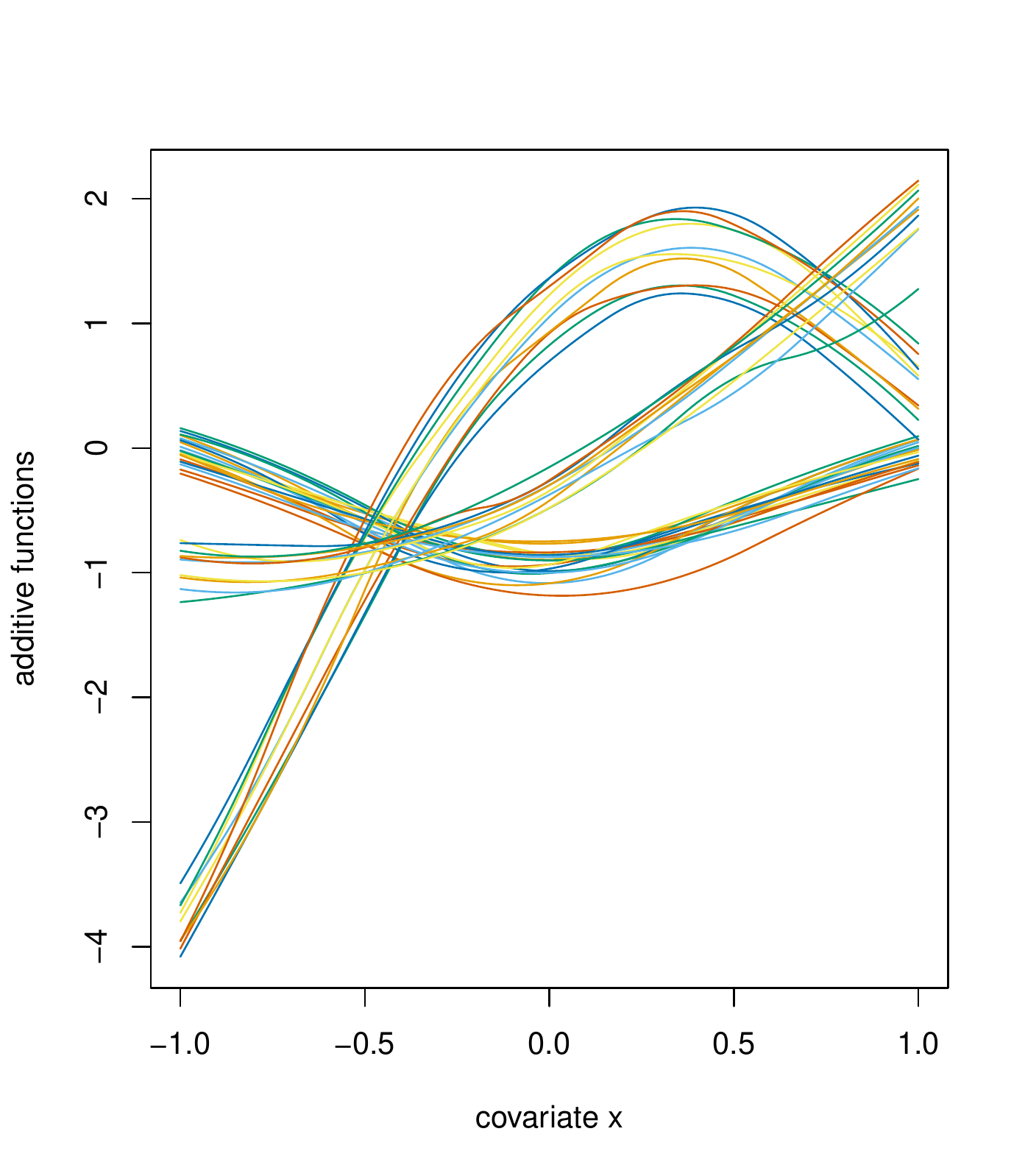}
  \caption{Additive functions}
\end{subfigure}%
\begin{subfigure}{.33\textwidth}
  \centering
  \includegraphics[width=1.0\textwidth]{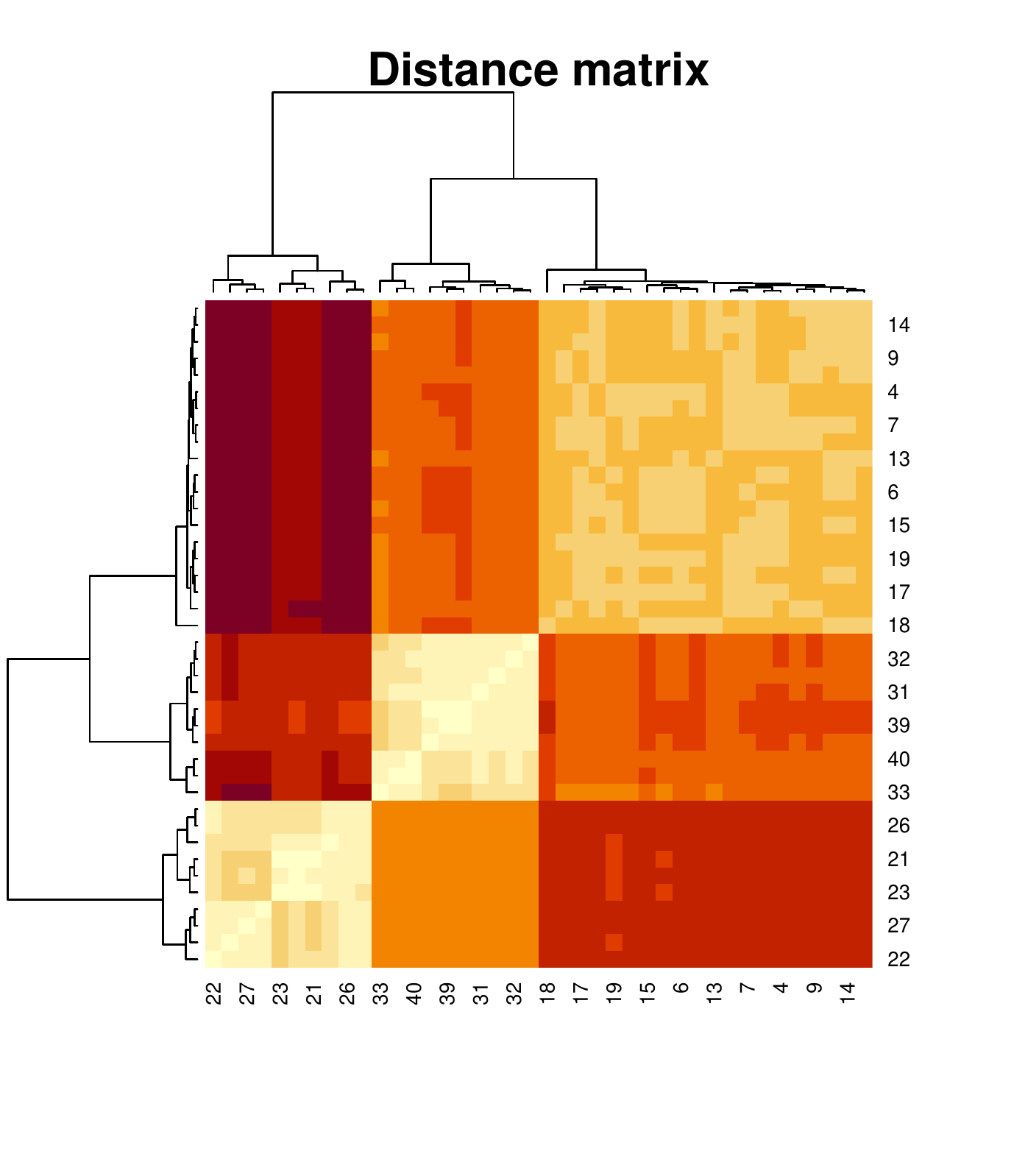}
  \caption{Distance matrix}
\end{subfigure}
\begin{subfigure}{.33\textwidth}
  \centering
  \includegraphics[width=1.0\textwidth]{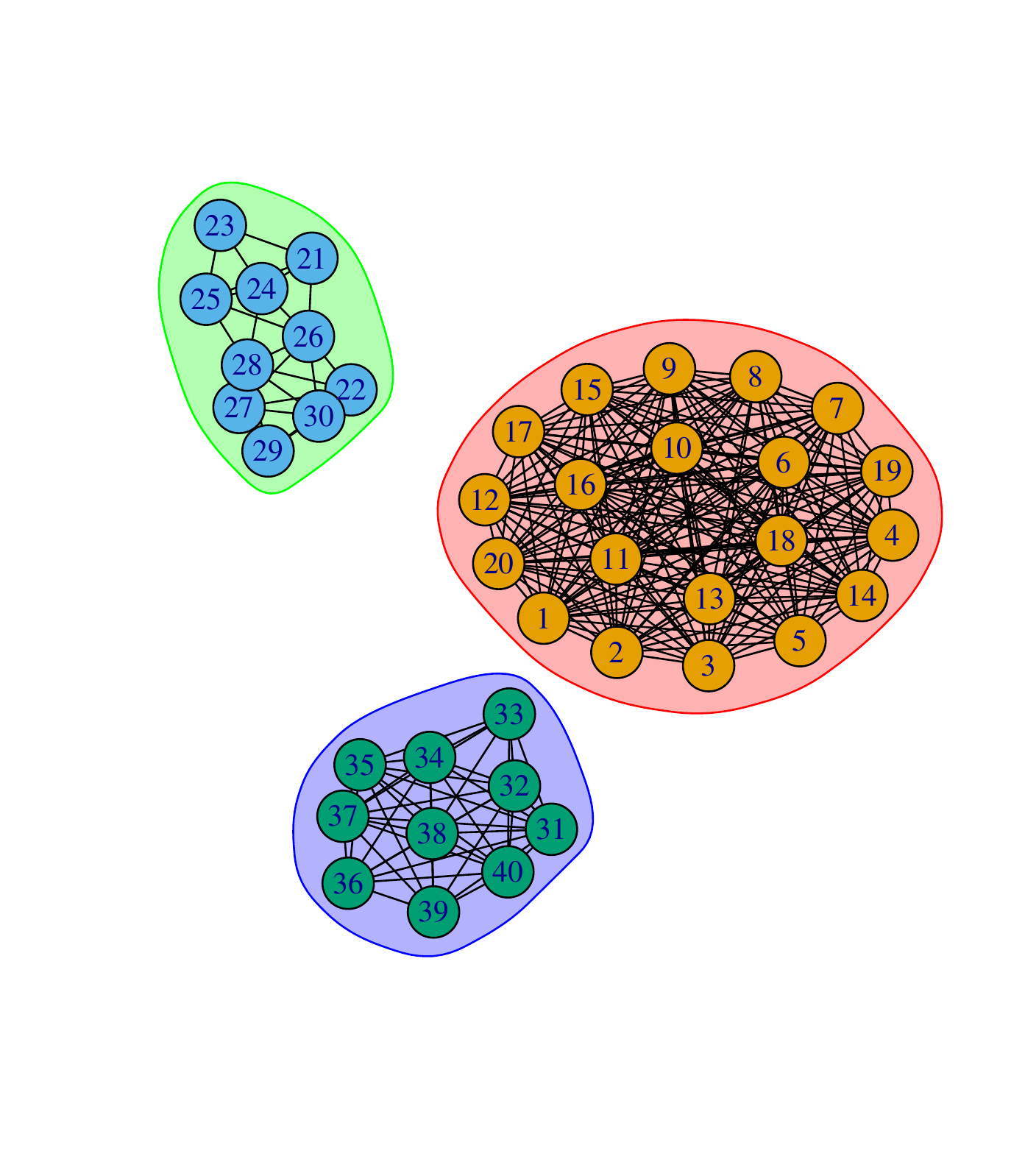}
  \caption{Group structure}
\end{subfigure}
\caption{One typical run of structure identification for $g_{i,j}(\cdot)$. (a) The estimates, $\hat{g}_{i,j}(\cdot)$, of specific additive functions for each subject $i$.  (b) The distance matrix constructed by the $L_{2}$ distance.  (c) The identified structure, where the numbers ($1,\ldots,40$) denote the combinations of $(i,j)$.}
\label{fig:flow2}
\end{figure}

Based on the initial estimation, we can identify the latent structure using the procedure in Section~\ref{hom0}. Note that in our simulation, the true partition for both $a_{i,j}(\cdot)$ and $g_{i,j}(\cdot)$ has three clusters (see Figures \ref{fig:initial_alpha} and \ref{fig:initial_beta}). Figures~\ref{fig:flow1} and~\ref{fig:flow2} present one simulation result of this step, which is aligned with our simulation setting. For example, in Figure~\ref{fig:flow1}, the left-hand side shows the initial estimates for $a_{i,j}(\cdot)$, the middle is the $L_{2}$ distance matrix calculated based on $a_{i,j}(\cdot)$ and the right-hand side is the result of the structure identification. The cluster membership is denoted by the colours.

The mean and standard deviation of the NMIs for $a_{i,j}(\cdot)$ and $g_{i,j}(\cdot)$ are shown in Table~\ref{Table:nmi}. We can observe that as $T$ gets larger, the performance becomes better. This makes sense since the initial estimates improve as the sample size $T$ increases.

\begin{table}[!ht] \renewcommand{\arraystretch}{1.2}
\caption{The mean and standard deviation of the NMIs for $\calD_H$ and $\Delta_m$ based on 100 simulations with different sample size}
\label{Table:nmi}
\begin{center}
{\footnotesize
\begin{tabular}{ccccccc}
\hline\hline
 & $n/T$ &  $T=100$  &  $T=200$  &  $T=400$     \\
\hline
\multirow{2}*{$\calD_H$}
& $n=20$ &0.8807(0.1406) & 0.9672(0.0596) &0.9953(0.0296) \\

&$n=40$ &0.8755(0.1113) &  0.9720(0.0581) & 0.9948(0.0244)\\

\hline
\multirow{2}*{$\Delta_m$}
& $n=20$ &0.9669(0.0573) & 0.9920(0.0456) &0.9987(0.0125) \\

&$n=40$ &0.8824(0.1198) &  0.9954(0.0181) &0.9961(0.0170) \\

\hline
\hline
\end{tabular}
}
\end{center}
\end{table}

\subsection{Final estimation}

\begin{figure}[!ht]
\centering
\includegraphics[width=1.0\textwidth,height=6cm]{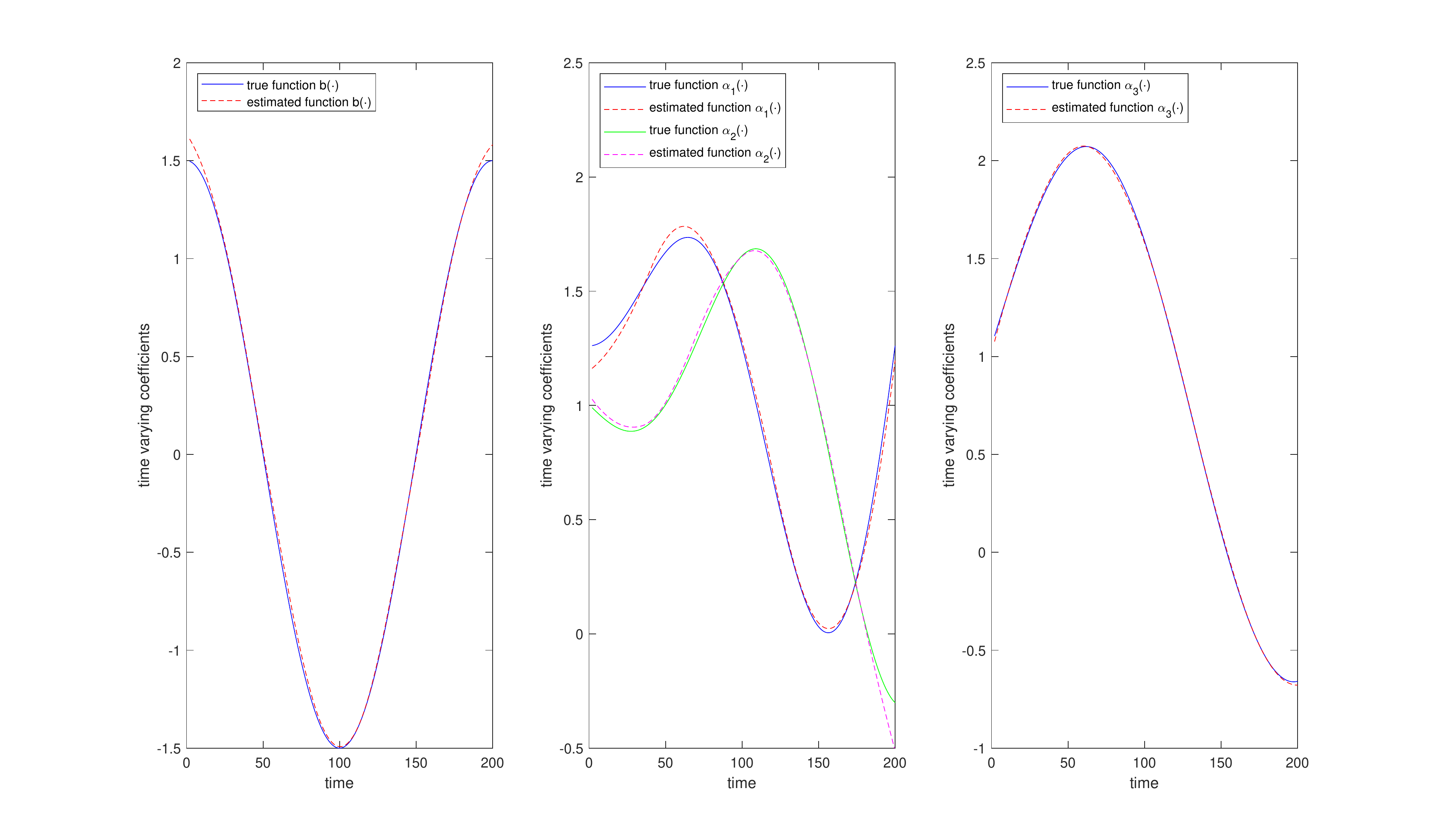}
\caption{Median estimated varying-coefficient functions $\alpha_{k}(\cdot)$ based on 100 simulations when $n=20,T=200$.}
\label{fig:alpha}
\end{figure}

\begin{figure}[!ht]
\centering
\includegraphics[width=1.0\textwidth,height=6cm]{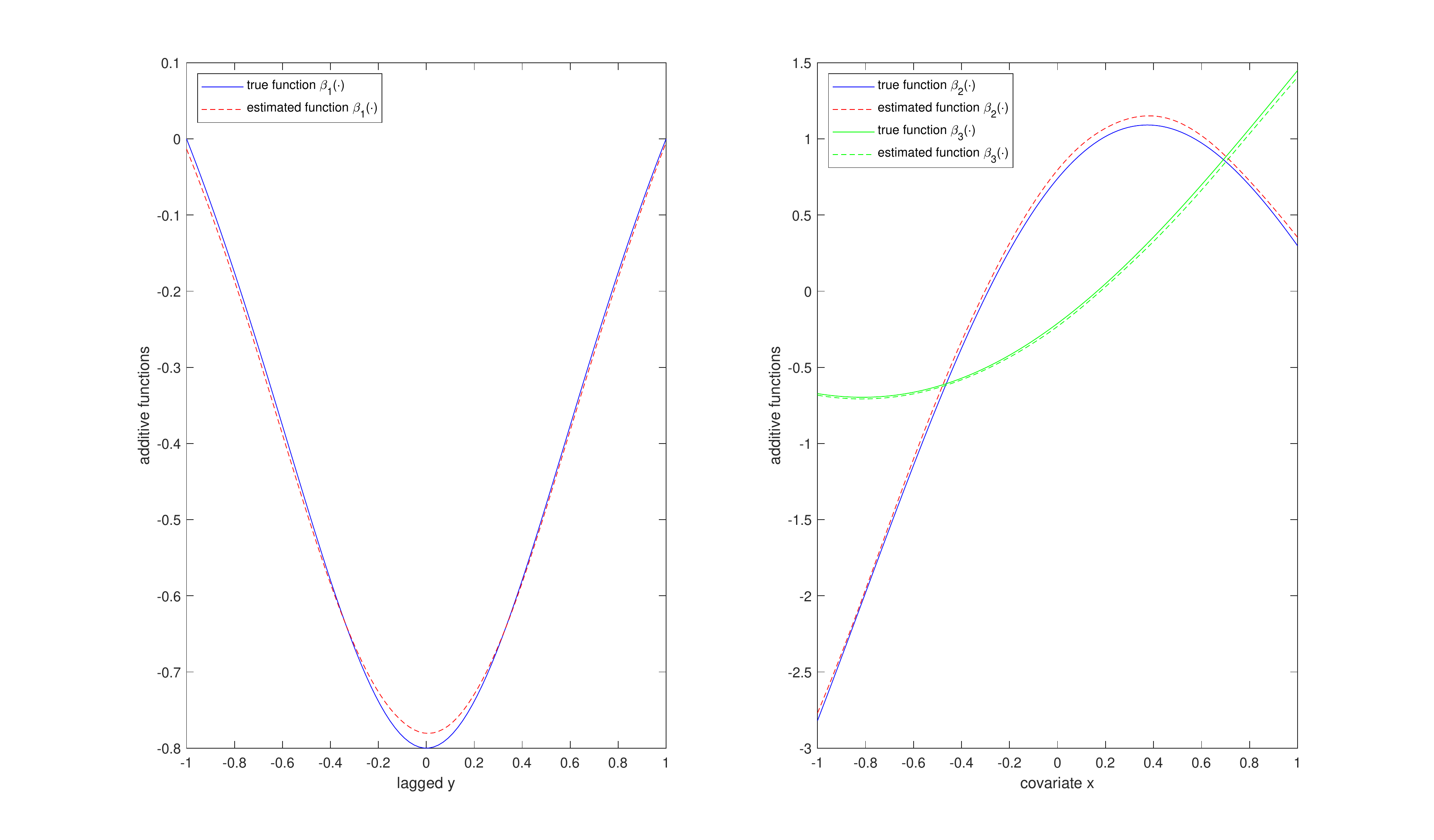}
\caption{Median estimated additive functions $\beta_{k}(\cdot)$ based on 100 simulations when $n=20,T=200$.}
\label{fig:beta}
\end{figure}

We have illustrated the performance of the structure identification.  We now assume that
the true cluster structure is known in this step and adopt the procedure in
Section~\ref{fin0} to estimate the cluster-specific functions.  The final
estimates are shown in Figures~\ref{fig:alpha} and~\ref{fig:beta}.  In
particular, we compare the estimation performance by plotting the final
estimates and the true functions in the same figure. Obviously, the final
estimates are all close to the true functions.

Finally, in Table~\ref{Table:mse}, we compare the performance of correct fitting with overfitting and underfitting. We can observe that among these three methods, correct fitting provides the best performance with the lowest MISE. For correct fitting and overfitting, the overall estimation accuracy for all the unknown functions almost improves as $T$ increases, which is somewhat expected, since as $T$ increases, the initial estimates are better and the subsequent procedures rely more on the initial estimates. Note that the functions estimated by underfitting are not consistent, which agrees with Theorem~\ref{thm:2}.

Note that overfitting and underfitting, which either ignore or mistakenly specify the structure, perform worse than correct fitting, which illustrates the importance of incorporating the flexible and parsimonious structure in clustered data analysis.

\begin{table}[!ht] \renewcommand{\arraystretch}{1.2}
\caption{The mean and standard deviation of MISE for the estimated functions based on 100 simulations when $n=20$.}
\label{Table:mse}
\begin{center}
{\footnotesize
\begin{tabular}{ccccccccccc}
\hline\hline
 \text{Function} & Method & $T=100$  &  $T=200$  &  $T=400$    \\
\hline
\multirow{3}*{$b(\cdot)$}

&\textit{Correct fitting} &0.0458(0.0144) & 0.0341(0.0141) &0.0273(0.0113) \\

&\textit{Overfitting}    &0.1898(0.0220) & 0.1372(0.0194) &0.1342(0.0186) \\

&\textit{Underfitting}   &0.1258(0.0177) & 0.1363(0.0146) &0.1521(0.0192) \\
\hline
\multirow{3}*{$a_{i,1}(\cdot)$}

&\textit{Correct fitting} &0.0796(0.0234) & 0.0389(0.0093) &0.0304(0.0064) \\

&\textit{Overfitting}    &0.2280(0.0523) & 0.1011(0.0109) &0.0783(0.0143) \\

&\textit{Underfitting}   &0.3915(0.0097) & 0.3970(0.0062) &0.4027(0.0099) \\
\hline
\multirow{3}*{$a_{i,2}(\cdot)$}

&\textit{Correct fitting} &0.0493(0.0090) & 0.0106(0.0025) &0.0063(0.0017) \\

&\textit{Overfitting}    &0.1281(0.0196) & 0.0415(0.0038) &0.0264(0.0024) \\

&\textit{Underfitting}   &0.1885(0.0071) & 0.0902(0.0041) &0.0713(0.0108) \\
\hline
\multirow{3}*{$g_{i,1}(\cdot)$}

&\textit{Correct fitting} &0.0422(0.0102) & 0.0159(0.0048) &0.0167(0.0039) \\

&\textit{Overfitting}    &0.1119(0.0121) & 0.0643(0.0047) &0.0642(0.0051) \\

&\textit{Underfitting}   &0.1584(0.0136) & 0.2242(0.0153) &0.3124(0.0197) \\
\hline
\multirow{3}*{$g_{i,2}(\cdot)$}

&\textit{Correct fitting} &0.0934(0.0068) & 0.0387(0.0033) &0.0405(0.0013) \\

&\textit{Overfitting}    &0.2003(0.0161) & 0.1426(0.0024) &0.0919(0.0016) \\

&\textit{Underfitting}   &0.5974(0.0008) & 0.6119(0.0006) &0.6163(0.0008) \\
\hline
\hline
\end{tabular}
}
\end{center}
\end{table}

\section{Analysis of Covid-19 data from China}
\label{real0}

With the severity of the new coronavirus disease (Covid-19) outbreak, much literature focused on the the prediction of growth trajectories, such as \citep[and references therein]{tang2021interplay,liu2021panel,li2021will}. Now we apply our method to analyse the Covid-19 data from a different perspective, mentioned in Section~\ref{intr0}. For each day, we collected the cumulative number of confirmed cases $\left(Z_{i, t}\right)$, the number of people travelling from Wuhan to other provinces (\url{https://qianxi.baidu.com/}) and the maximum daily temperature (\url{http://www.weather.com.cn}).

The infection ratio is the response variable. It is denoted by $y_{i, t}$ and is measured by $\log \left(Z_{i, t}\right)- \log \left(Z_{i, t-1}\right)$.\footnote{The time series plot of $y_{i,t}$ for 29 provinces in China from 23 January to 8 April is presented in the supplementary material.} To explore how the influential factors contribute to the infection ratio, we focus on first-order lagged dependent variable $y_{i,t-1}$ and two covariates: (1) the proportion of the population of Wuhan travelling to the $i$th province on day $t-14$, denoted by $x_{i, t,1}$; (2) the maximum daily
temperature in the $i$th province on day $t$, denoted by $x_{i,t,2}$.  We first
standardise $y_{i,t-1}$ and the covariates $x_{i, t,1}$ and $x_{i,t,2}$ to $[0,1]$, then apply the model (\ref{sta1}), with $p=1$, $u_{i,t} = t/T$, $i=1, \ldots, 29$ and $t=1, \dots, 77$, to fit the data.  The proposed
estimation procedure is implemented to estimate the unknown functions in the model.

\begin{figure}[!ht]
\centering
\begin{subfigure}{.49\textwidth}
  \centering
  \includegraphics[width=1.0\textwidth]{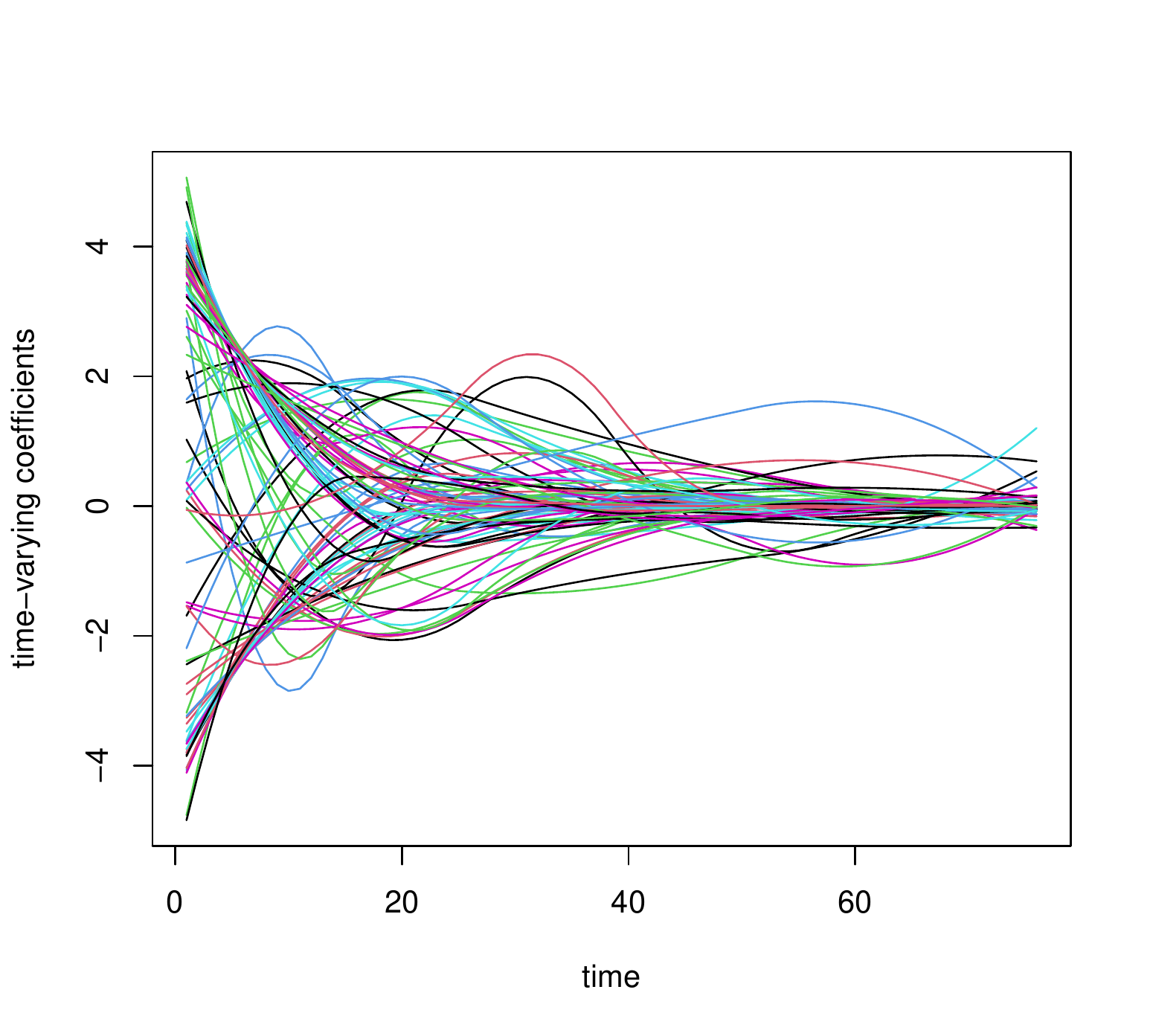}
  \caption{Estimated time-varying coefficients $\hat{a}_{i,j}(\cdot)$}
\end{subfigure}%
\begin{subfigure}{.49\textwidth}
  \centering
  \includegraphics[width=1.0\textwidth]{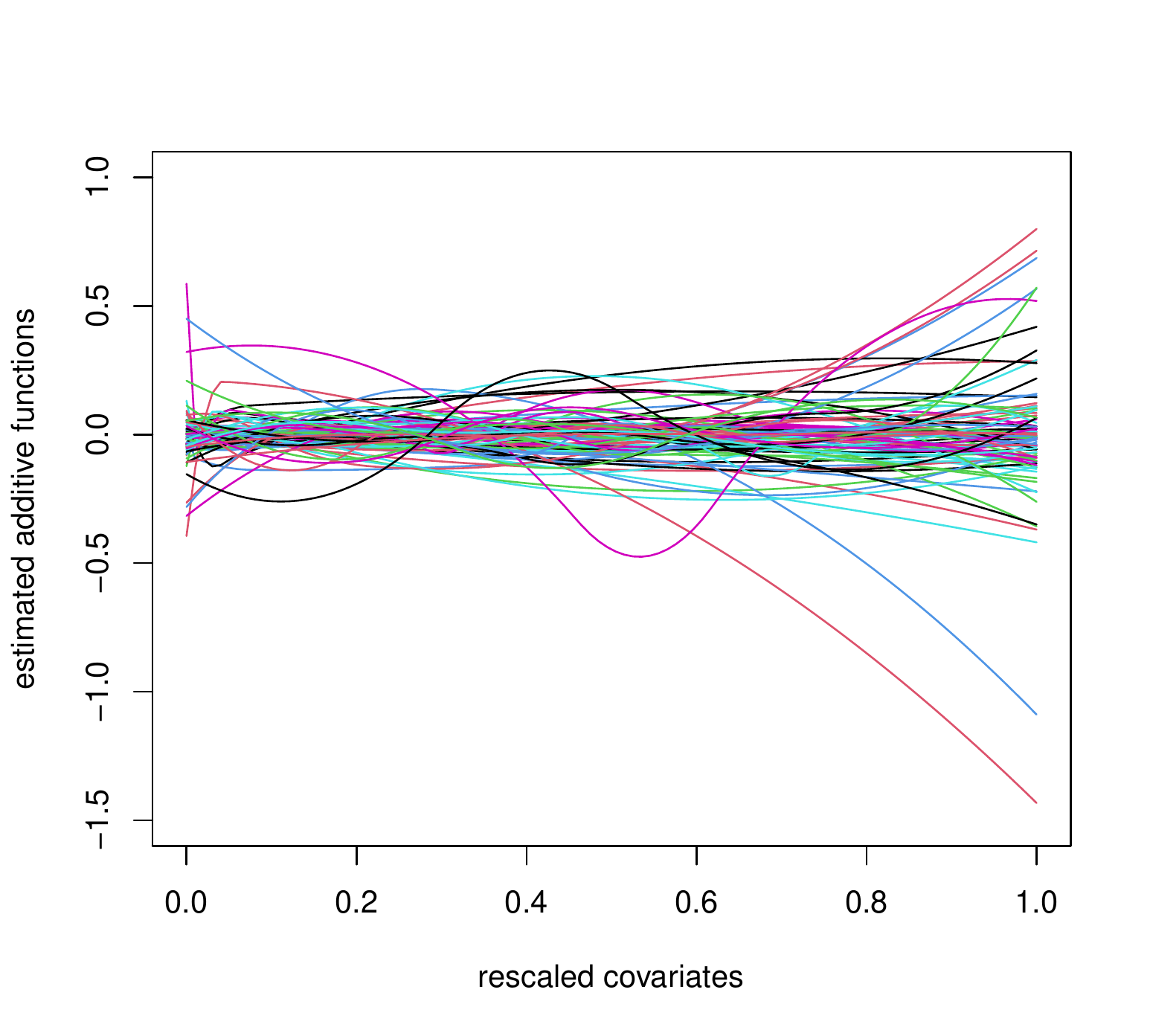}
  \caption{Estimated additive functions $\hat{g}_{i,j}(\cdot)$}
\end{subfigure}
\caption{(a) Initial estimated time-varying coefficient functions
$\hat{a}_{i,j}(\cdot)$ for $i=1,\ldots,29; j=1,2,3$. (b) Initial estimated additive
functions $\hat{g}_{i,j}(\cdot)$ for $i=1,\ldots,29; j=1,2,3$.
}
\label{fig:covid1}
\end{figure}

Specifically, we first apply the proposed method to estimate the
province-specific coefficients, the obtained initial estimates of time-varying
coefficients and additive functions are presented in Figure~\ref{fig:covid1}.
We can see that both $\hat{a}_{i,j}(\cdot)$ and $\hat{g}_{i,j}(\cdot)$ have
some homogeneous structures, which verifies the necessity of considering a
latent structure model to characterise this homogeneity.  A question naturally
arises: How many clusters are there and which provinces share similar impacts?

In the structure identification step, we calculate the $L_{2}$ distance between different functions, and identify the latent structure based on the proposed method in Section~\ref{hom0}. The identified cluster memberships are presented in Tables \ref{Table:a} and \ref{Table:b}. To save space, we put the visualization maps of these cluster memberships in the supplementary material.
\begin{table}[!ht]
 \renewcommand{\arraystretch}{1.0}
\caption{Identified cluster membership for time-varying coefficients $\hat{a}_{i,j}(\cdot)$.}
\label{Table:a}
\begin{center}
{\footnotesize
\begin{tabular}{c|c|c|c}
\hline
Cluster & $\hat{a}_{i,1}(\cdot)$ & $\hat{a}_{i,2}(\cdot)$ & $\hat{a}_{i,3}(\cdot)$\\
\hline
\multirow{6}*{Cluster 1} & Anhui, Gansu,         & Anhui, Guizhou,            & Beijing, Chongqing, Gansu,\\
                         & Guangdong, Guizhou,   & Heilongjiang, Henan,       & Hainan, Heilongjiang, Hunan\\
                         & Henan, Jiangsu,       & Hunan, Jilin,              & Inner mongolia, Jiangxi, Liaoning\\
                         & Jiangsu, Jilin,       &  Shanxi, Tianjin           & Qinghai, Shaanxi, Shanghai\\
                         & Liaoning, Ningxia,                            &    & Shanxi, Sichuan, Tianjin,\\
                         & Qinghai, Xinjiang, Yunnan                     &    & Yunnan, Zhejiang\\
\hline
\multirow{4}*{Cluster 2} & Beijing, Fujian, Guangxi,    & Guangxi, Hainan,  & Anhui, Fujian, Guangdong, \\
                         & Hainan, Hebei, Heilongjiang, & Inner mongolia, Jiangsu,        & Guangxi, Guizhou, Hebei,\\
                         & Inner mongolia, Shanghai, Shanxi,& Jiangxi, Ningxia,  & Jiangsu, Xinjiang\\
                         & Tianjin, Zhejiang &Shaanxi, Xinjiang, Zhejiang &\\
\hline
Cluster 3 & Chongqing, Shaanxi, Sichuan & Gansu, Guangdong, Liaoning  & Henan, Shandong\\

\hline
\multirow{3}*{Cluster 4} & \multirow{3}*{Shandong} & Beijing, Chongqing, Fujian & \multirow{3}*{Jilin, Ningxia} \\
                         &          & Hebei, Qinghai, Shandong & \\
                         &          & Sichuan, Yunnan & \\
\hline
Cluster 5 & Hunan &-&- \\
\hline
Cluster 6 & - & Shanghai & -\\
\hline
\end{tabular}
}
\end{center}
\end{table}

\begin{figure}[!ht]
\centering
\includegraphics[width=1.0\textwidth,height=10cm]{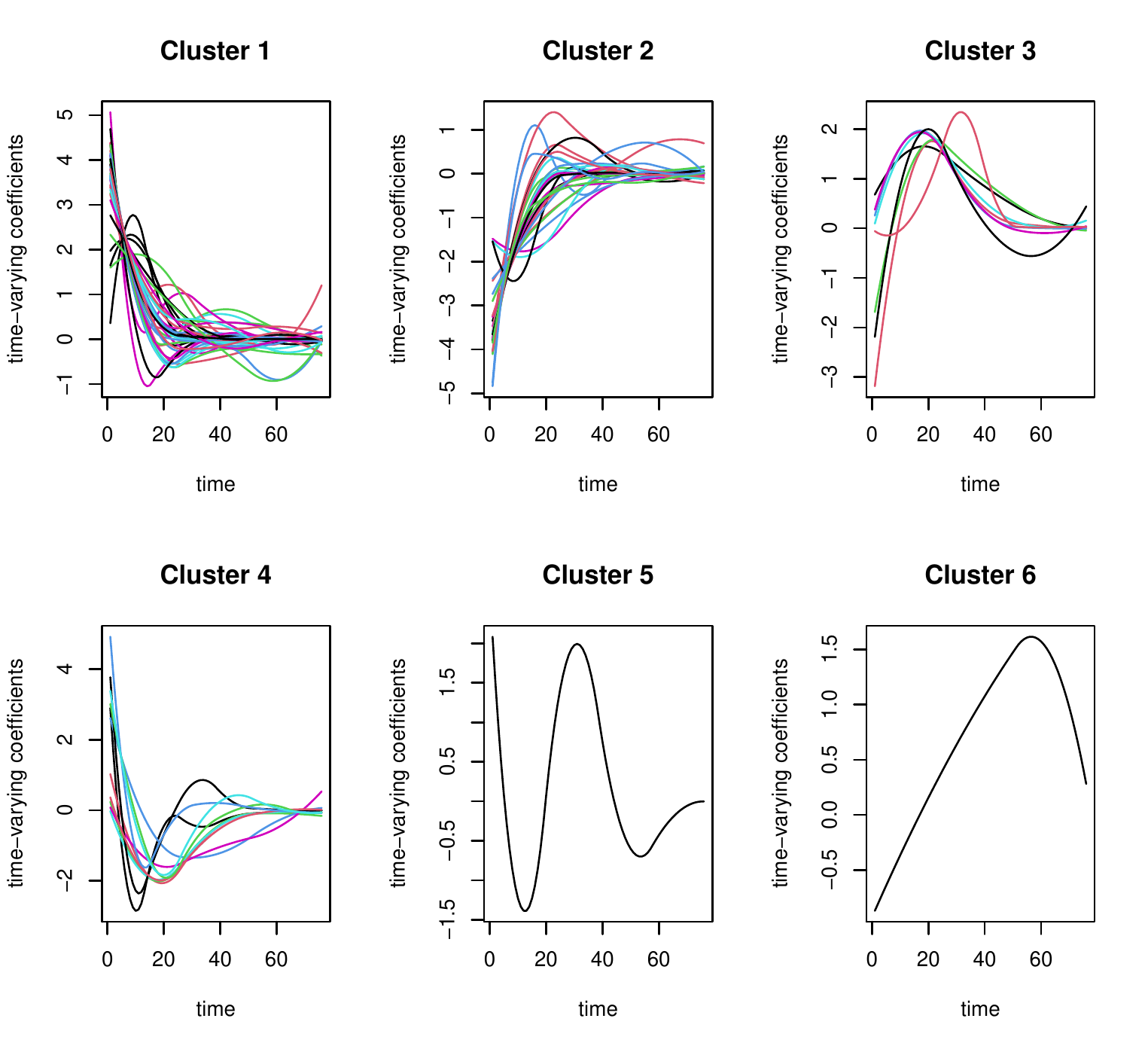}
\caption{Identified latent structure for time varying coefficient functions $\hat{a}_{i,j}(\cdot)$, where $i=1,\ldots,29, j=1,2,3$.}
\label{fig:group1}
\end{figure}

\begin{figure}[!ht]
\centering
\includegraphics[width=1.0\textwidth,height=10cm]{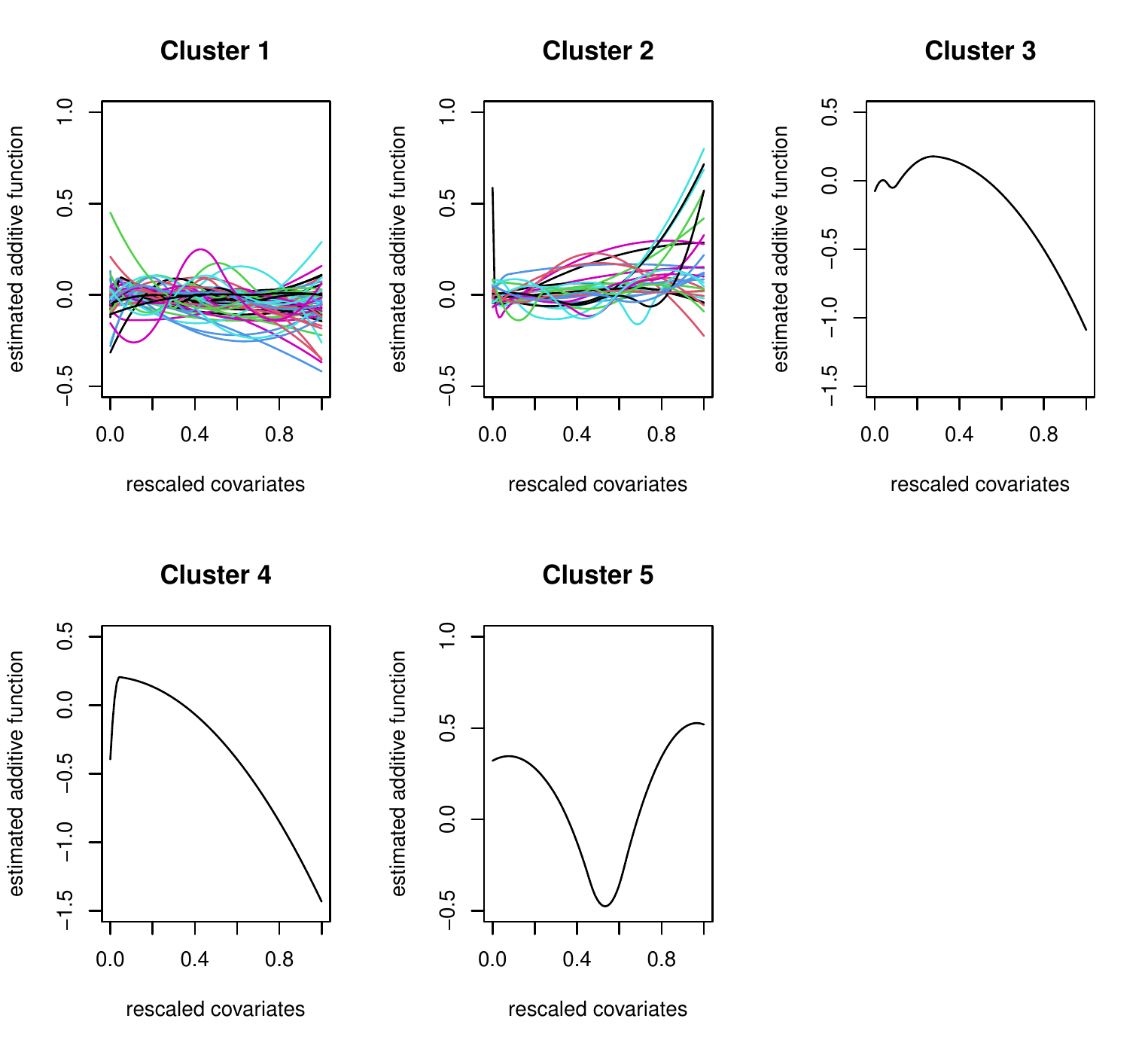}
\caption{Identified latent structure for additive functions $\hat{g}_{i,j}(\cdot)$, where $i=1,\ldots,29, j=1,2,3$. }
\label{fig:group2}
\end{figure}

\begin{table}[!ht]
 \renewcommand{\arraystretch}{1.0}
\caption{Identified cluster membership for additive functions $\hat{g}_{i,j}(\cdot)$.}
\label{Table:b}
\begin{center}
{\footnotesize
\begin{tabular}{c|c|c|c}
\hline
Cluster & $\hat{g}_{i,1}(\cdot)$ & $\hat{g}_{i,2}(\cdot)$ & $\hat{g}_{i,3}(\cdot)$\\
\hline
\multirow{8}*{Cluster 1} & Anhui, Chongqing, Fujian,         & Fujian, Guangdong,    & Anhui, Beijing, Chongqing,\\
                         & Gansu, Guangxi, Guizhou,         & Guangxi, Hebei,       & Gansu, Guangdong, Hebei\\
                         & Hainan, Hebei,                   & Hunan, Jiangxi,       & Henan, Hunan, Inner mongolia,\\
                         & Heilongjiang, Inner mongolia,    & Jilin, Liaoning,      & Jilin, Liaoning, Ningxia,\\
                         & Jiangsu, Jiangxi, Jilin,         & Ningxia, Qinghai,     & Qinghai, Shandong, Shanghai,\\
                         & Ningxia, Qinghai, Shaanxi,       & Shanghai, Shanxi,     & Shanxi, Sichuan, Tianjin\\
                         & Shanghai, Sichuan, Tianjin,      &  Yunnan                & Xinjiang, Yunnan\\
                         & Xinjiang, Yunnan, Zhejiang        &&\\
\hline
\multirow{5}*{Cluster 2} & Beijing, Guangdong,     & Anhui, Beijing, Chongqing,      & Fujian, Guangxi, \\
                         & Henan, Hunan,     & Gansu, Guizhou, Hainan,         & Hainan, Heilongjiang,\\
                         & Liaoning, Shandong,                       & Henan, Inner mongolia, Jiangsu, & Jiangsu, Jiangxi,\\
                         & Shanxi                               & Shaanxi, Shandong, Sichuan      & Shaanxi, Zhejiang\\
                         &                               & Tianjin, Zhejiang & \\
\hline
Cluster 3 & - & Heilongjiang  & -\\

\hline
Cluster 4 & - & Xinjiang & - \\
\hline
Cluster 5 & - &-&Guizhou \\
\hline
\end{tabular}
}
\end{center}
\end{table}
For each identified cluster, we plot the initial estimates of time-varying
coefficients or additive functions in that cluster, and present them in
Figures~\ref{fig:group1} and~\ref{fig:group2}.  From Figures~\ref{fig:group1}
and~\ref{fig:group2}, we can roughly see that the coefficients
$\hat{a}_{i,j}(\cdot)$ and $\hat{g}_{i,j}(\cdot)$ belonging to the same cluster
have similar patterns, which illustrates that our proposed structure identification
approach performs well.

\begin{figure}[!ht]
\centering
\begin{subfigure}{.33\textwidth}
  \centering
  \includegraphics[width=1.0\textwidth,height=8cm]{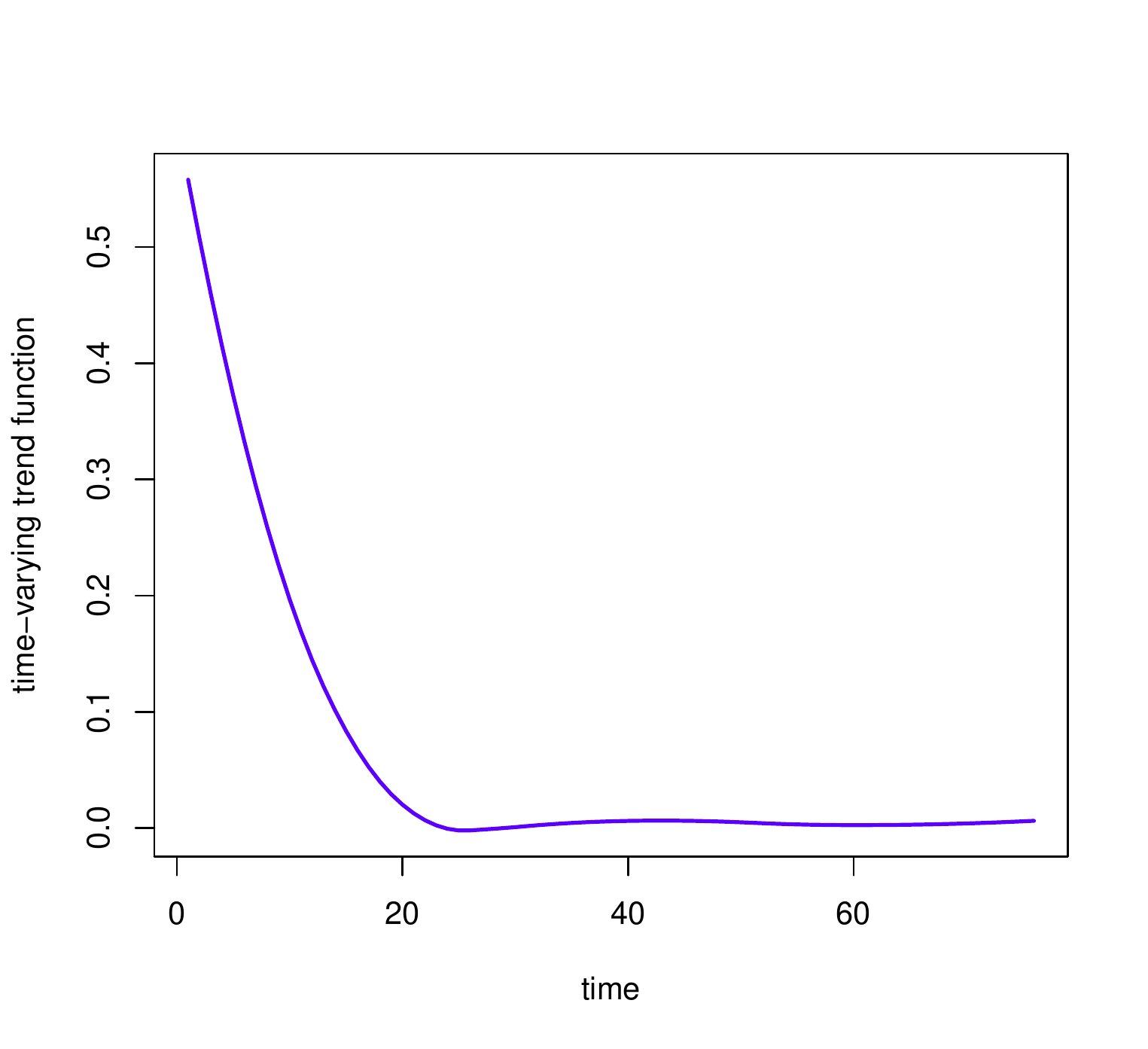}
\end{subfigure}%
\begin{subfigure}{.33\textwidth}
  \centering
  \includegraphics[width=1.0\textwidth,height=8cm]{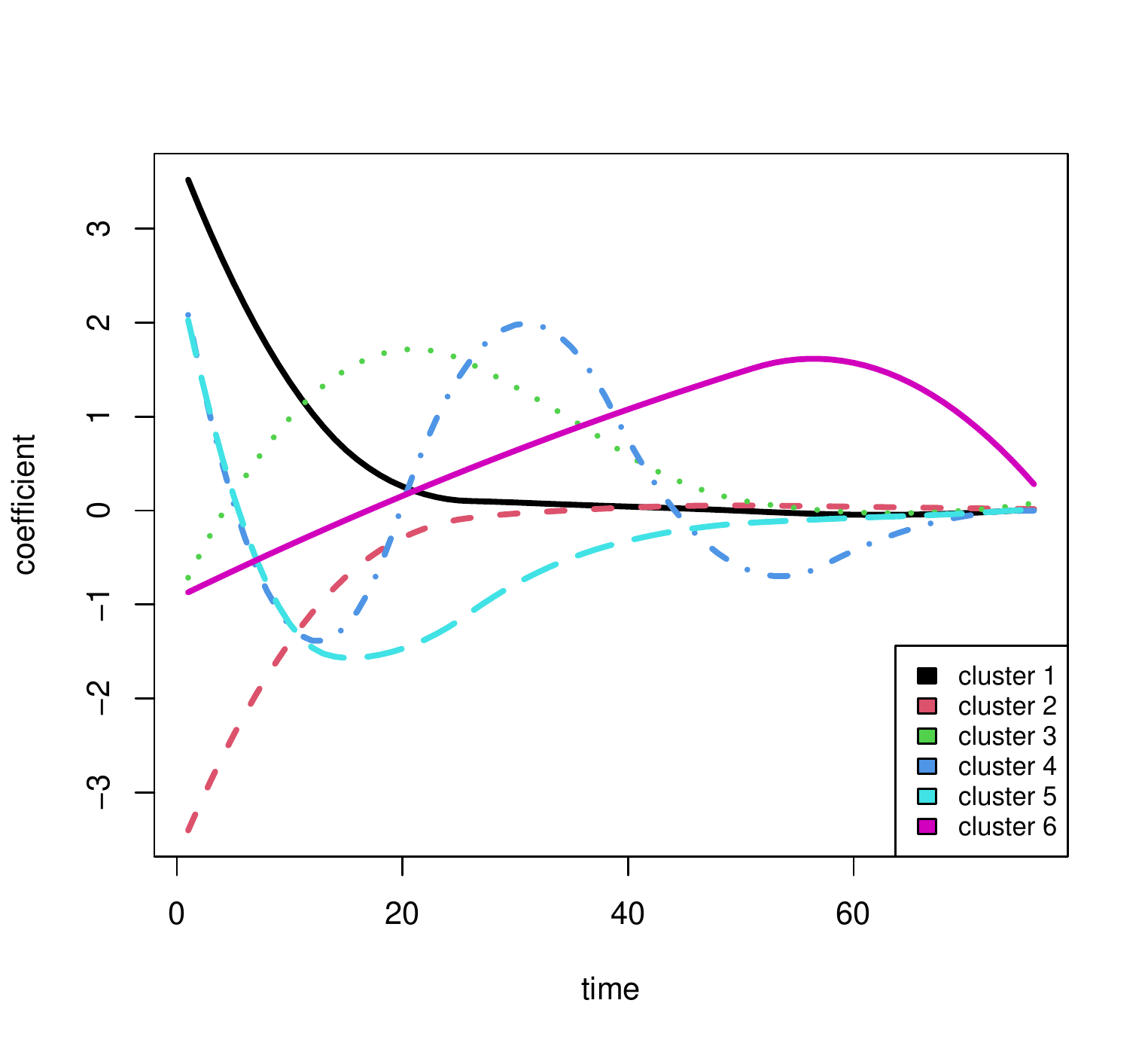}
\end{subfigure}%
\begin{subfigure}{.33\textwidth}
  \centering
  \includegraphics[width=1.0\textwidth,height=8cm]{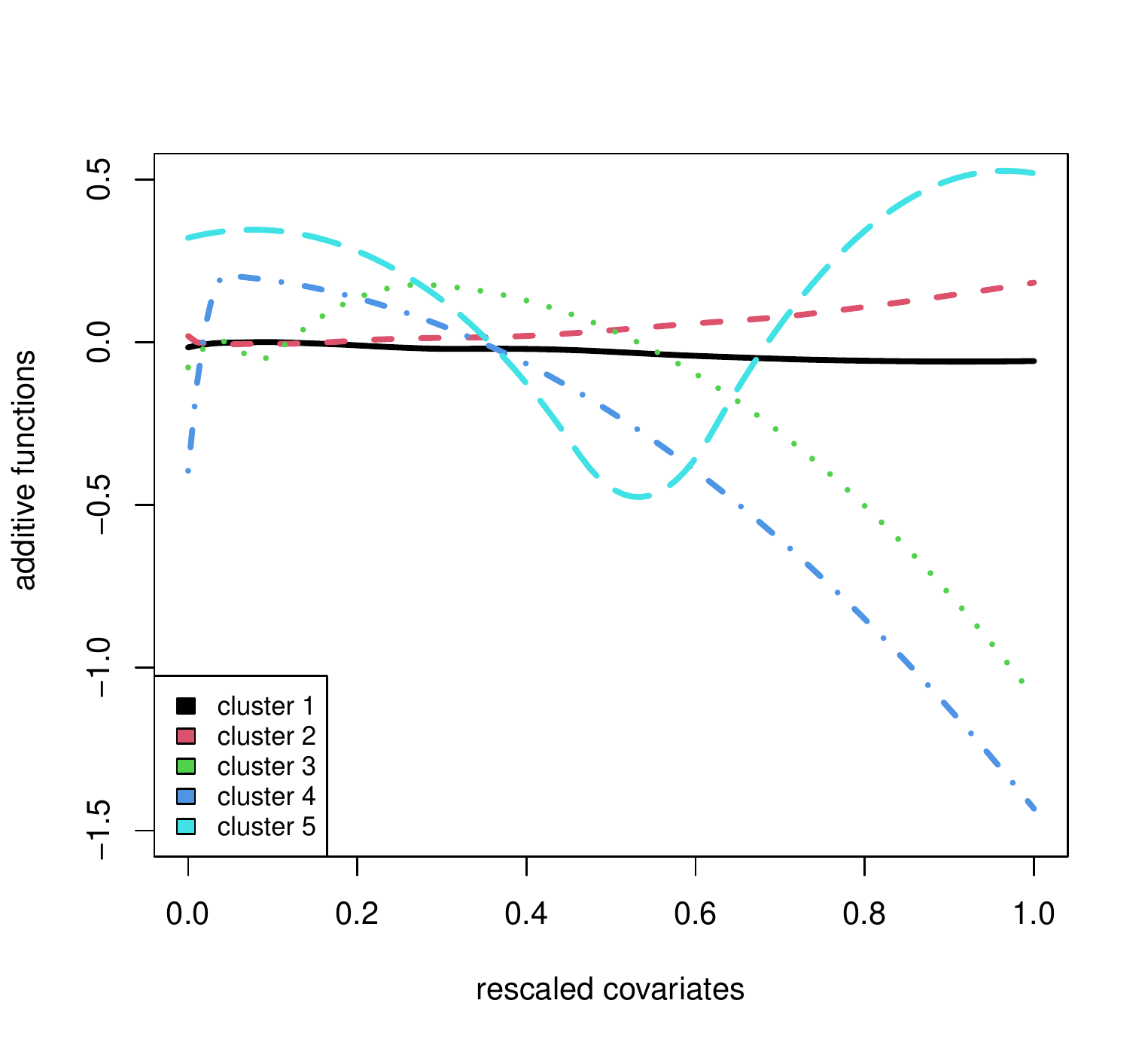}
\end{subfigure}
\caption{Final estimates of unknown functions. Left: estimated trend function $b(\cdot)$. Middle: estimated cluster-specific time-varying coefficient functions $\alpha_{k}(\cdot), k=1,\ \cdots,6$. Right: estimated cluster-specific additive functions $\beta_{k}(\cdot), k=1,\ \cdots,5$.}
\label{fig:final1}
\end{figure}

Finally, based on the identified structure, we estimate $b(\cdot)$, $\alpha_{k}(\cdot)$ and $\beta_{k}(\cdot)$ according to the procedure proposed in Section~\ref{fin0}. The results are presented in Figure~\ref{fig:final1}, where the left panel of Figure~\ref{fig:final1} denotes the estimated trend function. From the middle panel of Figure~\ref{fig:final1}, we can conclude that the previous infection rate, maximum daily temperature and the number of people who travelled from Wuhan to the province have dynamic impacts on the infection ratio and the impacts vary over different provinces. The provinces which share the same impacts are depicted in Table~\ref{Table:a}. From the right panel of Figure~\ref{fig:final1}, we can also find that the contribution of covariates to response is through transformed covariates and the contributions vary over different provinces. The cluster-specific functions and its corresponding functional changes are depicted in Table~\ref{Table:group}.

\begin{table} \renewcommand{\arraystretch}{1.2}
\caption{Summary descriptions for the identified cluster-specific functions}
\label{Table:group}
\begin{center}
{\footnotesize
\begin{tabular}{c|c|c}
\hline\hline
 \text{Function} & Cluster & Cluster-specific function description \\
\hline
\multirow{1}*{$b(\cdot)$}

&Cluster 1& The coefficients decreases from positive to zero gradually. \\

\hline
\multirow{6}*{$\alpha_{k}(\cdot)$}
& Cluster 1  & The coefficients decreases from positive to zero gradually.\\
& Cluster 2  & The coefficients increases from negative to zero gradually.\\
& Cluster 3   & The coefficients first increases from negative to positive, then decreases to zero. \\
& Cluster 4  & The coefficients show a ``W" shape.\\
& Cluster 5  & The coefficients first decreases from positive to negative, then increases to zero.\\
& Cluster 6   & The coefficients show an inverted ``U'' shape.\\
\hline
\multirow{5}*{$\beta_{k}(\cdot)$}

& Cluster 1  & The function is a constant function with values zero.\\
& Cluster 2  & The function is a monotonic increasing function with positive values.\\
& Cluster 3  & The function is a mixture of two quadratic functions.\\
& Cluster 4  & The function a monotonic decreasing function with negative values.\\
& Cluster 5  & The function is a quadratic function, the values first decreases and then increases.\\
\hline
\hline
\end{tabular}
}
\end{center}
\end{table}

Furthermore, we examine the out of sample prediction performance between the proposed model (\ref{sta1}) and the varying coefficient additive model (VCAM) without the lagged terms $y_{i,t-1}$. For a given fitted model, let $e_{i,j} = (y_{i,T-j} - \hat{y}_{i,T-j})^2$ be the squared prediction error for the infection rate of the $i$th province on the $(T-j)$th day in the sample, based on the fitted model using all the observations before the $(T-j)$th day. Following \cite{li2018factor} lead, we construct a cross-validation prediction error for the last 14 days as
\[
\mathrm{PE} = \frac{1}{14 \times 29}\sum^{29}\limits_{i=1}\sum^{14}\limits_{j=1}e_{i,j}.
\]
The PE for our proposed model (\ref{sta1}) is 0.0053, and 0.0669 for VCAM, which verifies our proposed model is better than the VCAM in terms of out-of-sample prediction.

\section{Concluding remarks}
\label{sec:conc}
This paper has presented a flexible and parsimonious modelling mechanism for identifying and estimating latent structure in a class of semiparametric clustered data models in which the varying coefficients and additive functions are heterogeneous across clusters but homogeneous within a cluster and the cluster membership is unknown. In particular, we have considered identifying the hidden structure by a distance-based approach and estimating the cluster-specific functions with a three-stage method. We also established the asymptotic properties for the estimators obtained by overfitting, underfitting and correct fitting. The results of the simulations and an analysis of real data were provided to illustrate the finite-sample performance of the proposed method.

\bibliographystyle{apalike}
\bibliography{reference}
\end{document}